\documentclass[twocolumn]{aastex631}

\newcommand\jwst{JWST}
\newcommand\alma{ALMA}
\newcommand\cigale{{\sc cigale}}
\newcommand\prospector{{\sc prospector}}

\newcommand\target{SPT0418$-$47}
\newcommand\main{SPT0418A}
\newcommand\companion{SPT0418B}
\newcommand\nircam{NIRCam}
\newcommand\miri{MIRI}

\usepackage{xspace}
\newcommand{\cii}{[C {\sc II}]\xspace}
\newcommand{\um}{\ensuremath{\mu\rm{m}}\xspace}
\newcommand{\kms}{\ensuremath{\rm{km\,s}^{-1}}\xspace}

\begin{document}
\journalinfo{Submitted to AAS Journals June 23, 2023}
\title{TEMPLATES: Characterization of a Merger in the Dusty Lensing SPT0418-47 System}

\correspondingauthor{Jared Cathey}
\email{jaredcathey@ufl.edu}

\author[0000-0002-4657-7679]{Jared Cathey}
\affiliation{Department of Astronomy, University of Florida, 211 Bryant Space Sciences Center, Gainesville, FL 32611 USA}
\author[0000-0002-0933-8601]{Anthony H. Gonzalez}
\affiliation{Department of Astronomy, University of Florida, 211 Bryant Space Sciences Center, Gainesville, FL 32611 USA}
\author[0000-0003-4422-8595]{Sidney Lower}
\affiliation{Department of Astronomy, University of Florida, 211 Bryant Space Sciences Center, Gainesville, FL 32611 USA}
\author[0000-0001-7946-557X]{Kedar A. Phadke}
\affiliation{Department of Astronomy, University of Illinois, 1002 West Green St., Urbana, IL 61801, USA}
\affiliation{Center for AstroPhysical Surveys, National Center for Supercomputing Applications, 1205 West Clark Street,
Urbana, IL 61801, USA}
\author[0000-0003-3256-5615]{Justin Spilker}
\affiliation{Department of Physics and Astronomy and George P. and Cynthia Woods Mitchell Institute for Fundamental
Physics and Astronomy, Texas A\&M University, 4242 TAMU, College Station, TX 77843-4242, USA}

\author[0000-0002-6290-3198]{Manuel Aravena}
\affiliation{Núcleo de Astronomía, Facultad de Ingeniería y Ciencias, Universidad Diego Portales, Av. Ejercito 441, Santiago, Chile}
\author[0000-0003-1074-4807]{Matthew Bayliss}
\affiliation{Department of Physics, University of Cincinnati, Cincinnati, OH 45221, USA}
\author[0000-0002-3272-7568]{Jack E. Birkin}
\affiliation{Department of Physics and Astronomy and George P. and Cynthia Woods Mitchell Institute for Fundamental
Physics and Astronomy, Texas A\&M University, 4242 TAMU, College Station, TX 77843-4242, USA}
\author[0000-0003-3195-5507]{Simon Birrer}
\affiliation{Department of Physics and Astronomy, Stony Brook University, Stony Brook, NY 11794, USA}
\author[0000-0002-8487-3153]{Scott Chapman}
\affiliation{Department of Physics and Atmospheric Science, Dalhousie University, Halifax, NS, B3H 4R2, Canada}\affiliation{NRC Herzberg Astronomy and Astrophysics, 5071 West Saanich Rd, Victoria, BC, V9E 2E7, Canada}\affiliation{Department of Physics and Astronomy, University of British Columbia, Vancouver, BC, V6T1Z1, Canada}\affiliation{Eureka Scientific Inc, Oakland, CA 94602, USA}
\author[0000-0003-2200-5606]{H\r{a}kon Dahle}
\affiliation{Institute of Theoretical Astrophysics, University of Oslo, P.O. Box 1029, Blindern, NO-0315 Oslo, Norway}
\author[0000-0003-4073-3236]{Christopher C. Hayward}
\affiliation{Center for Computational Astrophysics, Flatiron Institute, 162 5th Avenue, 10010, New York, NY, USA}
\author[0000-0002-8669-5733]{Yashar Hezaveh}
\affiliation{Center for Computational Astrophysics, Flatiron Institute, 162 5th Avenue, 10010, New York, NY, USA}
\affiliation{Department of Physics, Université de Montréal, Montréal, Canada}
\affiliation{Ciela - Montreal Institute for Astrophysical Data Analysis and Machine Learning, Montréal, Canada}
\affiliation{Mila - Quebec Artificial Intelligence Institute, Montréal, Canada}
\author{Ryley Hill}
\affiliation{Department of Physics and Astronomy, University of British Columbia, Vancouver, BC, V6T1Z1, Canada}
\author[0000-0001-6251-4988]{Taylor A. Hutchison}
\affiliation{Observational Cosmology Lab, Code 665, NASA Goddard Space Flight Center, 8800 Greenbelt Rd., Greenbelt, MD 20771, USA}\altaffiliation{NASA Postdoctoral Fellow}
\author[0000-0001-6505-0293]{Keunho J. Kim}
\affiliation{Department of Physics, University of Cincinnati, Cincinnati, OH 45221, USA}
\author[0000-0003-3266-2001]{Guillaume Mahler}
\affiliation{Centre for Extragalactic Astronomy, Durham University, South Road, Durham DH1 3LE, UK} 
\affiliation{Institute for Computational Cosmology, Durham University, South Road, Durham DH1 3LE, UK}
\author[0000-0002-2367-1080]{Daniel P. Marrone}
\affiliation{Steward Observatory, University of Arizona, 933 North Cherry Avenue, Tucson, AZ 85721, USA}
\author[0000-0002-7064-4309]{Desika Narayanan}
\affiliation{Department of Astronomy, University of Florida, Gainesville, FL USA 32611-2055}
\author[0000-0001-7548-0473]{Alexander Navarre}
\affiliation{Department of Physics, University of Cincinnati, Cincinnati, OH 45221, USA}
\author[0000-0001-7477-1586]{Cassie Reuter}
\affiliation{Department of Astronomy, University of Illinois, 1002 West Green St., Urbana, IL 61801, USA}
\author[0000-0002-7627-6551]{Jane R. Rigby}
\affiliation{Observational Cosmology Lab, Code 665, NASA Goddard Space Flight Center, Greenbelt, MD 20771}
\author[0000-0002-7559-0864]{Keren Sharon}
\affiliation{University of Michigan, Department of Astronomy, 1085 South University Avenue, Ann Arbor, MI 48109, USA}
\author[0000-0001-6629-0379]{Manuel Solimano}
\affiliation{Instituto de Estudios Astrof\'{\i}sicos, Facultad de Ingenier\'{\i}a y Ciencias, Universidad Diego Portales, Avenida Ej\'ercito Libertador 441, Santiago, Chile. [C\'odigo Postal 8370191]}
\author[0000-0002-3187-1648]{Nikolaus Sulzenauer}
\affiliation{Max-Planck-Institut für Radioastronomie, Auf dem Hügel 69, 53121 Bonn, Germany}
\author[0000-0001-7192-3871]{Joaquin Vieira}
\affiliation{Department of Astronomy, University of Illinois, 1002 West Green St., Urbana, IL 61801, USA}
\affiliation{Department of Physics, University of Illinois, 1110 West Green St., Urbana, IL 61801, USA}
\affiliation{Center for AstroPhysical Surveys, National Center for Supercomputing Applications, 1205 West Clark Street, Urbana, IL 61801, USA}
\author[0000-0001-7610-5544]{David Vizgan}
\affiliation{Department of Astronomy, University of Illinois, 1002 West Green St., Urbana, IL 61801, USA}


\begin{abstract}
We present \jwst\ and \alma\ results for the lensing system \target, which includes a strongly-lensed, dusty star-forming galaxy at redshift $z=4.225$ and an associated multiply-imaged companion. \jwst\ \nircam\ and \miri\ imaging observations presented in this paper were acquired as part of the Early Release Science program Targeting Extremely Magnified Panchromatic Lensed Arcs and Their Extended Star Formation (TEMPLATES). This data set provides robust, mutiwavelength detection of stellar light in both the main (\main) and companion (\companion) galaxies, while the \alma\ detection of \cii  emission confirms that \companion\ lies at the same redshift as \main. From a source plane reconstruction, we infer that the projected physical separation of the two galaxies is $4.42\pm 0.05$ kpc. We derive total magnifications of $\mu=29.5\pm1.2$ and $\mu=4.2\pm 0.9$ for \main\ and \companion, 
respectively. 
We use both \cigale\ and \prospector\ 
to derive stellar masses. 
The stellar mass ratio of \main\ and \companion\ is approximately 4 to 1 ($4.5\pm 1.0$ for \cigale\  and $4.2^{+1.9}_{-1.6}$ for \prospector).  
We also see evidence of extended structure associated with \main\ in the lensing reconstruction that is suggestive of a tidal feature. 
Interestingly, the star formation rates and stellar masses of both galaxies are consistent with the main sequence of star-forming galaxies at this epoch, indicating that this ongoing interaction has not noticeably elevated the star formation levels.
\end{abstract}


\section{Introduction} \label{sec:intro}
In the standard paradigm of galaxy formation, present-day spiral galaxies are products of hierarchical assembly, with the disks arising after the last major merger either as a byproduct of dissipative mergers 
\citep[e.g.][]{robertson2006ApJ...645..986R,governato2009MNRAS.398..312G} or from subsequent gas accretion and minor dissipative mergers \citep[e.g.][]{baugh1996MNRAS.283.1361B,steinmetz2002NewA....7..155S}. In this picture, dynamically cold disks become increasingly rare at higher redshifts \citep[e.g.][]{hopkins2010ApJ...715..202H}. Disks are expected to be highly turbulent and clumpy, driving significant outflows, 
and the ratio of rotational velocity to turbulent velocity, $V/\sigma$, should decrease with increasing redshift \citep{pillepich2019MNRAS.490.3196P}. 
Moreover, mergers can induce bursts of star formation, and hence can drive star formation at early times when mergers are expected to be more frequent \citep{sanders1988ApJ...325...74S, hopkins2008ApJS..175..356H,sotilloramos2022MNRAS.516.5404S}.
Studies of high-redshift submillimeter galaxies (SMGs) have supported a picture in which many of these strongly star-forming galaxies are major mergers \citep{engel2010ApJ...724..233E,alaghband2012MNRAS.424.2232A,2018Natur.553...51M,2019ApJ...870...80L,perry2022arXiv221008191P}. 
\begin{figure*}
    \centering
    \includegraphics[width=0.18\linewidth]{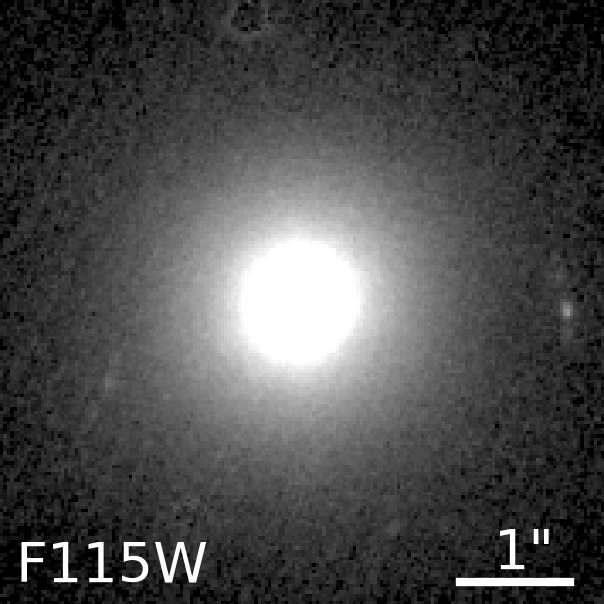}
    \includegraphics[width=0.18\linewidth]{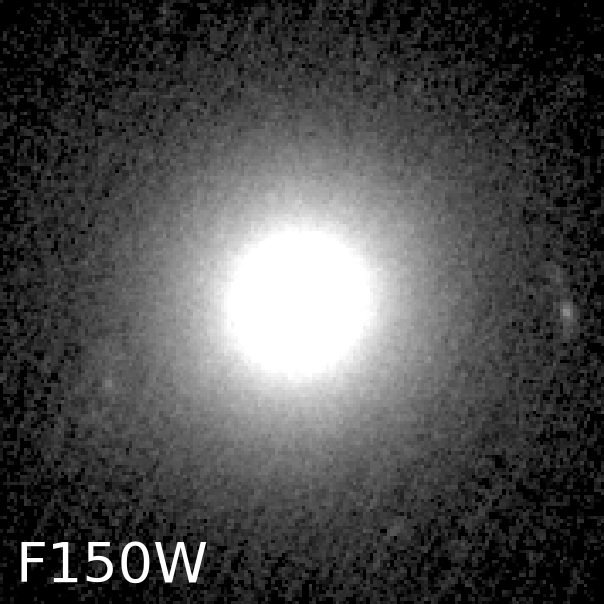}
    \includegraphics[width=0.18\linewidth]{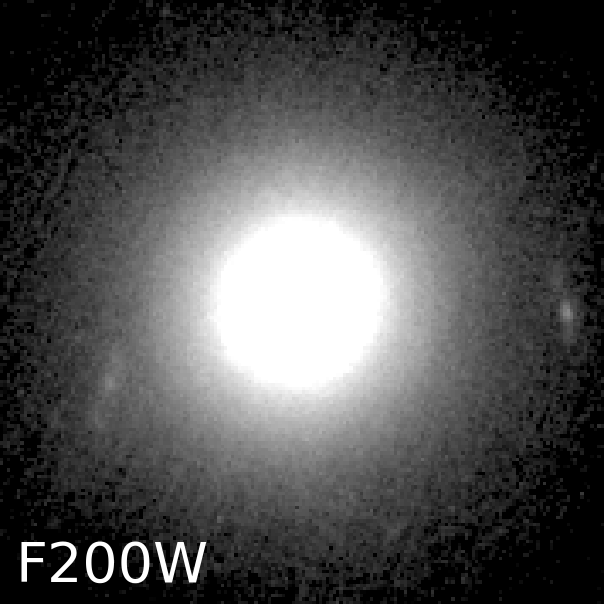}
    \includegraphics[width=0.18\linewidth]{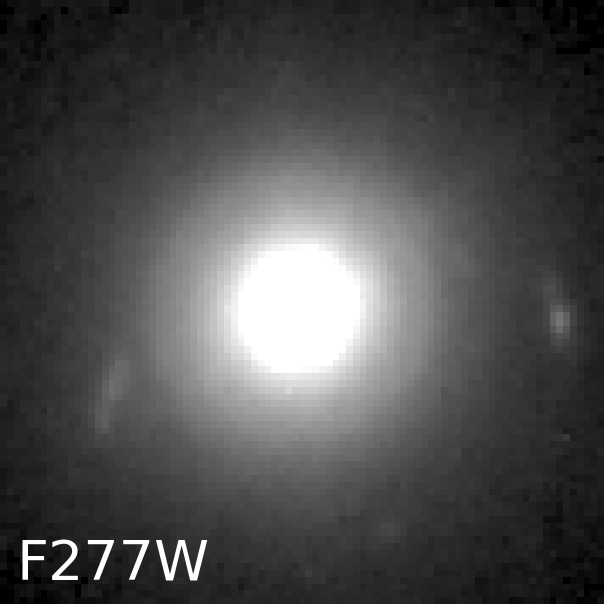}
    \includegraphics[width=0.18\linewidth]{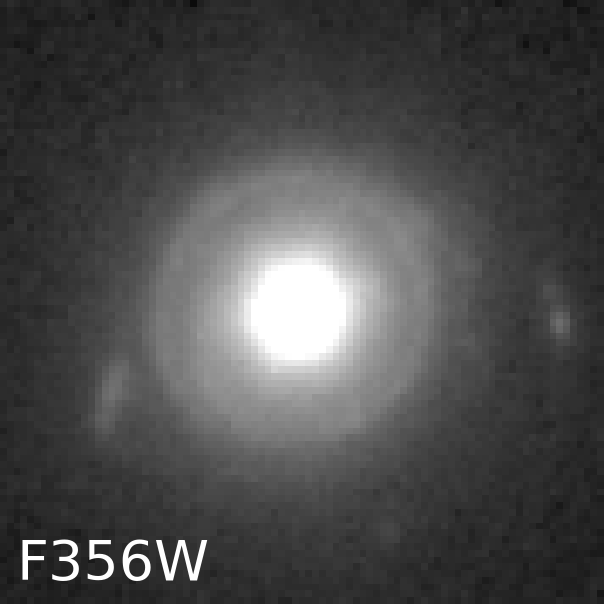}
    \includegraphics[width=0.18\linewidth]{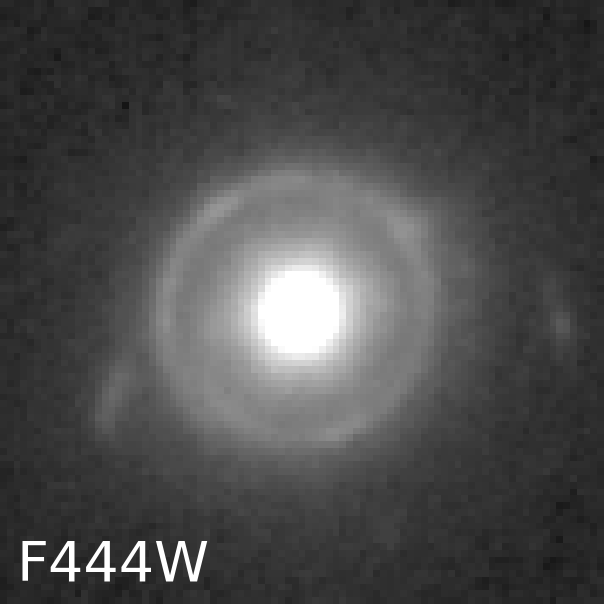}
    \includegraphics[width=0.18\linewidth]{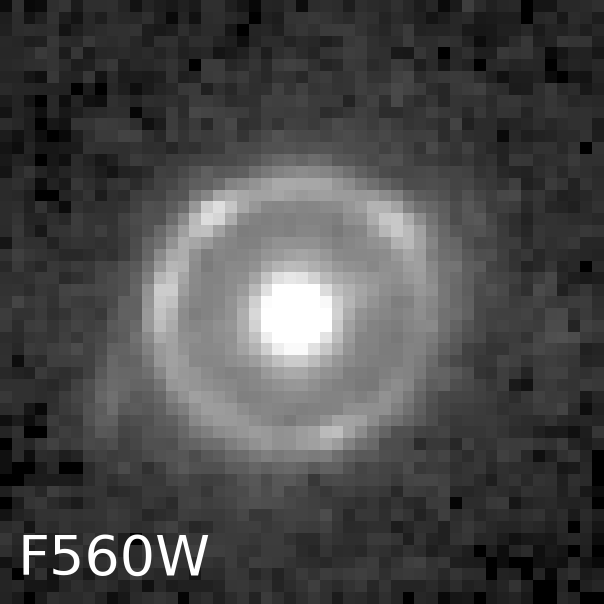}
    \includegraphics[width=0.18\linewidth]{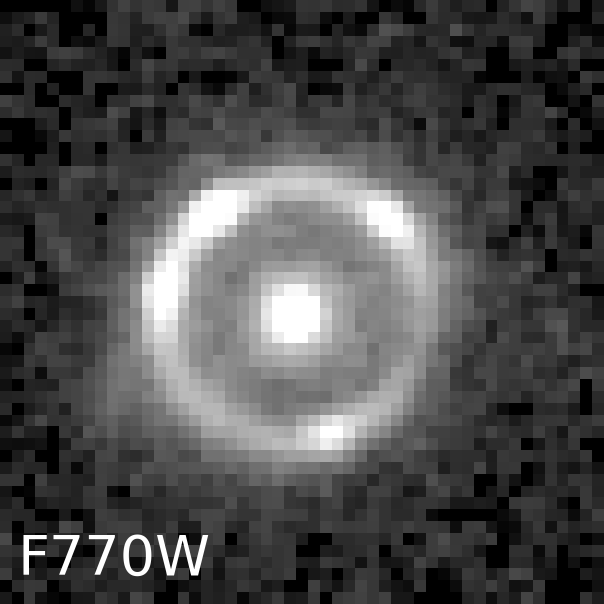}
    \includegraphics[width=0.18\linewidth]{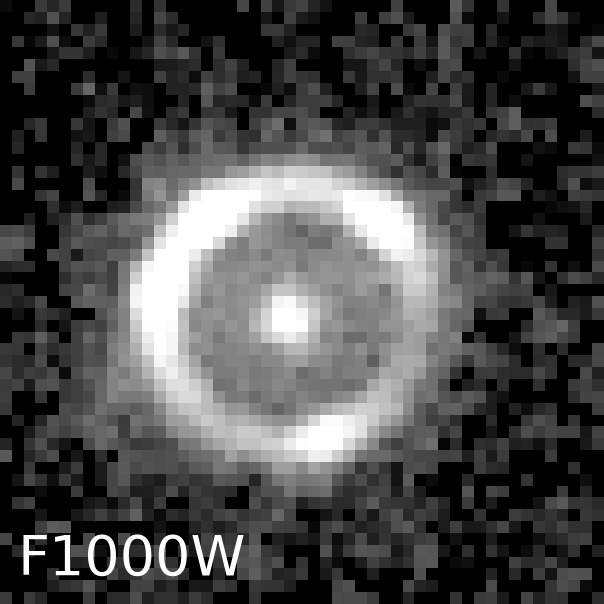}
    \includegraphics[width=0.18\linewidth]{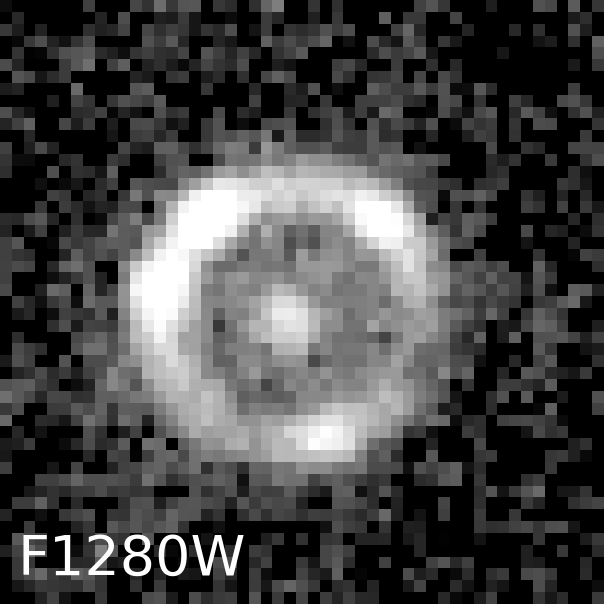}
    \includegraphics[width=0.18\linewidth]{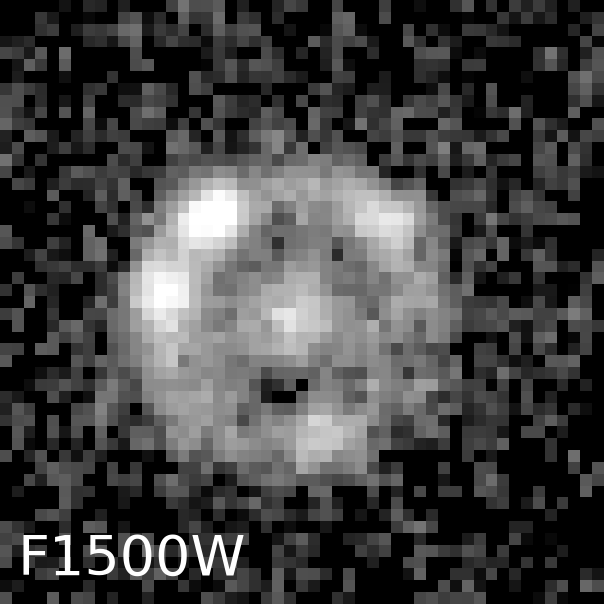}
    \includegraphics[width=0.18\linewidth]{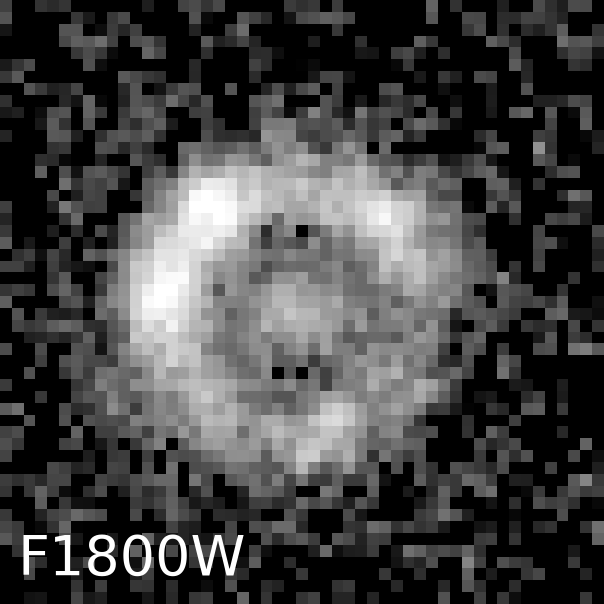}
    \includegraphics[width=0.18\linewidth]{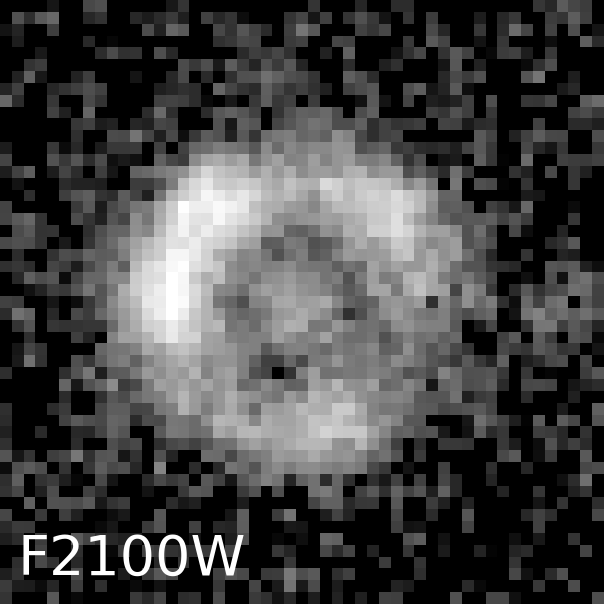}

    \caption{ Images of \target\ in all 13 NIRCam and MIRI passbands ranging from 1.15$\mu$m -- 21$\mu$m. These images illustrate the contrast as a function of wavelength between the foreground lens, \main, and \companion. Specific locations of \companion\ are shown in Figure \ref{fig:galfit}. The images are displayed with square-root scaling. The minimum value of the flux density scale is set to 0.1 MJy/sr below sky; the maximum in each frame is 2 MJy/sr above the sky level. A 1$^{\prime\prime}$ scale bar is included in the upper left panel.}   
    \label{fig:greyscale}
\end{figure*}

A counterpoint to this picture was presented by \citet{rizzo2020Natur.584..201R}, who found that the dusty, strongly star-forming galaxy \target\ at $z=4.225$ exhibits orderly disk rotation with $V/\sigma=9.7\pm0.4$ for the gas, based upon observations from the Atacama Large Millimeter Array (ALMA).
This ratio is a factor of $\sim3$ higher than predicted at this epoch \citep{pillepich2019MNRAS.490.3196P},
and the lack of dynamical signatures of a merger argued that the high star formation rate is driven purely by internal processes. Using ALMA, \citet{romanoliveira2023arXiv230203049R} have more recently found additional regularly rotating disks at $z\sim4.5$, and \citet{nelson2022arXiv220801630N} have uncovered with the James Webb Space Telescope (\jwst) a population of massive, dusty disks at $z=2-6$.

In this paper we re-examine \target\ using 
\jwst\ imaging and new ALMA observations. \target\ is one of four targets of the \jwst\ Early Release Science (ERS) program Targeting Extremely Magnified Panchromatic Lensed Arcs and their Extended Star formation (TEMPLATES, ERS Program 1355, PI: Rigby ; Co-PI: Vieira). The overall aim of TEMPLATES is to use multiple tracers to study the spatially-resolved star formation in four strongly lensed galaxies spanning a range of redshift and star formation rates. \target\ has the highest redshift of these four targets.

Originally discovered as a submillimeter source by the South Pole Telescope \citep[][]{vieira2013Natur.495..344V}, \target\ is known to have a high star formation rate 
($\sim280$ M$_\odot$ yr$^{-1}$) with  a dense, solar-metallicity interstellar medium (ISM)  from observations with  the Atacama Pathfinder EXperiment (APEX) and ALMA \citep{debreuck2019A&A...631A.167D}. \citet{spilker2020ApJ...905...86S} also observed an outflow of molecular gas in this galaxy  with $\dot M\sim 150$ M$_\odot$ yr$^{-1}$.

In this paper, we use NIRCam 
(Rieke, M., Kelly, D. M., Misselt, K., et al. 2023, submitted to PASP) and MIRI 
(Wright, G., Rieke, G. H., Glasse, A., et al. 2023, submitted to PASP) imaging to spatially resolve stellar emission in this system. 

We combine \jwst\ imaging and ALMA \cii data and investigate the environment of this system to discern whether this system is isolated or is dynamically interacting. As part of this analysis, we perform lens modeling to reconstruct the source plane image of \target.

We describe the data in \S \ref{sec:data}. In \S \ref{sec:companion} we describe the nearby companion, which was also reported by \citet{peng2022arXiv221016968P}. Hereafter we designate the companion \companion\ and the main galaxy \main. Next, in \S \ref{sec:merger} we discuss the lens modeling, dynamical state of the system, conduct Spectral Energy Distribution (SED) modeling, and estimate the stellar mass. 
In \S \ref{sec:conclusion} we have our concluding remarks. Throughout we assume a Planck cosmology \citep{planck2020A&A...641A...6P}: $H_0 = (67.4 \pm 0.5) \text{ km s}^{-1} \text{ Mpc}^{-1}$, $\Omega_{\text{m}} = 0.315 \pm 0.007$, $\Omega_\Lambda = 0.685 \pm 0.007$. 

\section{Data and Processing} \label{sec:data}
\subsection{\jwst}
 Instruments on board JWST have been shown to be working well, and in some cases performing better than expected \citep{Rigby_2023}.
Imaging observations of \target\ were taken by the NIRCam and MIRI instruments on 2022 August 11 and 2022 August 22, respectively, as part of the TEMPLATES program. 
Imaging was taken with filters spanning wavelengths from 1.15 $\mu$m to 21 $\mu$m, with the target centered in detector B4 for NIRCam. 
The filters and corresponding exposure times are given in the Appendix in Table \ref{tab:imaging}. 

Rigby et al. 2023 (in prep) describes in detail the data reduction process for TEMPLATES. Here we summarize the process for the subset of TEMPLATES data analyzed in this paper.
Starting with the Level 2A data products for NIRCam, we applied a custom de-striping algorithm to correct for $1/f$ noise and jumps between amplifiers. The de-striped images were then run through the JWST pipeline (Version 1.8.2)  using the CRDS context {\tt jwst\_0988.pmap}.

This version of the official \jwst\ calibration was the most up-to-date at the time of this analysis, and is consistent to within 3\% of the absolute flux calibration in
\citet{boyer2022}.

For the MIRI imaging, we used a four-point dither pattern optimized for extended sources. 
Uncalibrated images were processed through the JWST pipeline version 1.9.5dev using pmap {\tt jwst\_1062.pmap}. There are known striping issues for MIRI imaging, mainly arising from detector $1/f$ noise. 
We implemented de-striping for the current data by creating a detector template using the four dither positions and removing it from each exposure after stage 2 of the pipeline.\footnote{\url{https://github.com/STScI-MIRI/Imaging_ExampleNB/blob/main/helpers/miri_clean.py}} These de-striped stage 2 data products were processed through the stage 3 pipeline, and the images output from stage 3 were used for the analysis in this paper.

We realigned the MIRI and NIRCam imaging to a common frame to correct for residual astrometric offsets, and generated a simulated point spread function (PSF) for each filter for the date of observation using WebbPSF version 1.0.1.dev126+g6d83a9d \citep{perrin2012,perrin2014} given the measured wavefront of the telescope, which is measured every 2 days \citep{McElwain_2023}. Cutouts of \target\ in the F115W through F2100W filters are shown in Figure \ref{fig:greyscale}, illustrating the relative contrast between the lensing galaxy and the \target\ system as a function of wavelength.

\begin{figure*}[t]
\centering
    \includegraphics[width=0.32\textwidth]{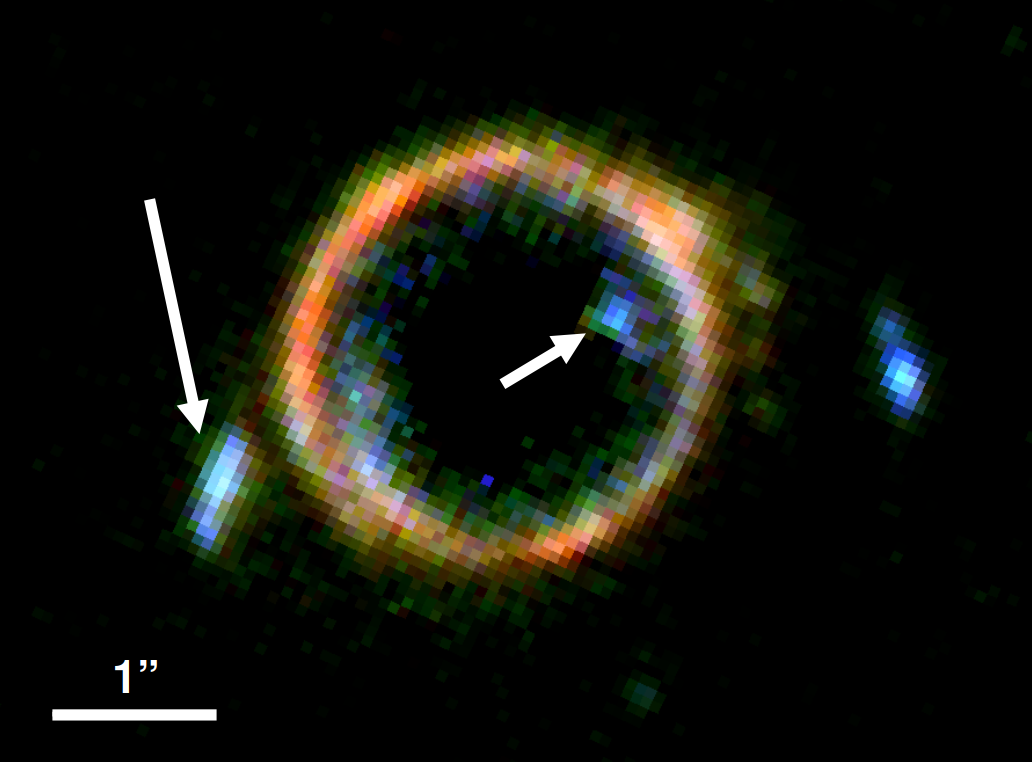}
    \hspace{0.005cm} 
    \includegraphics[width=0.32\textwidth]{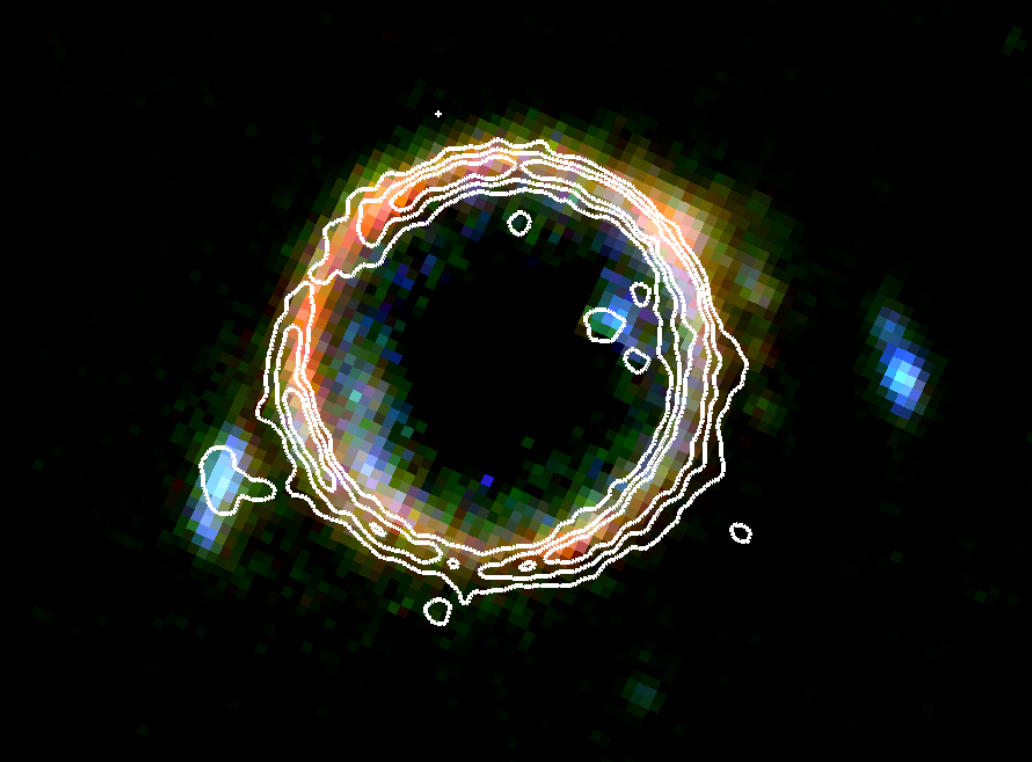}  
    \hspace{0.005cm}
    \includegraphics[width=0.32\textwidth]{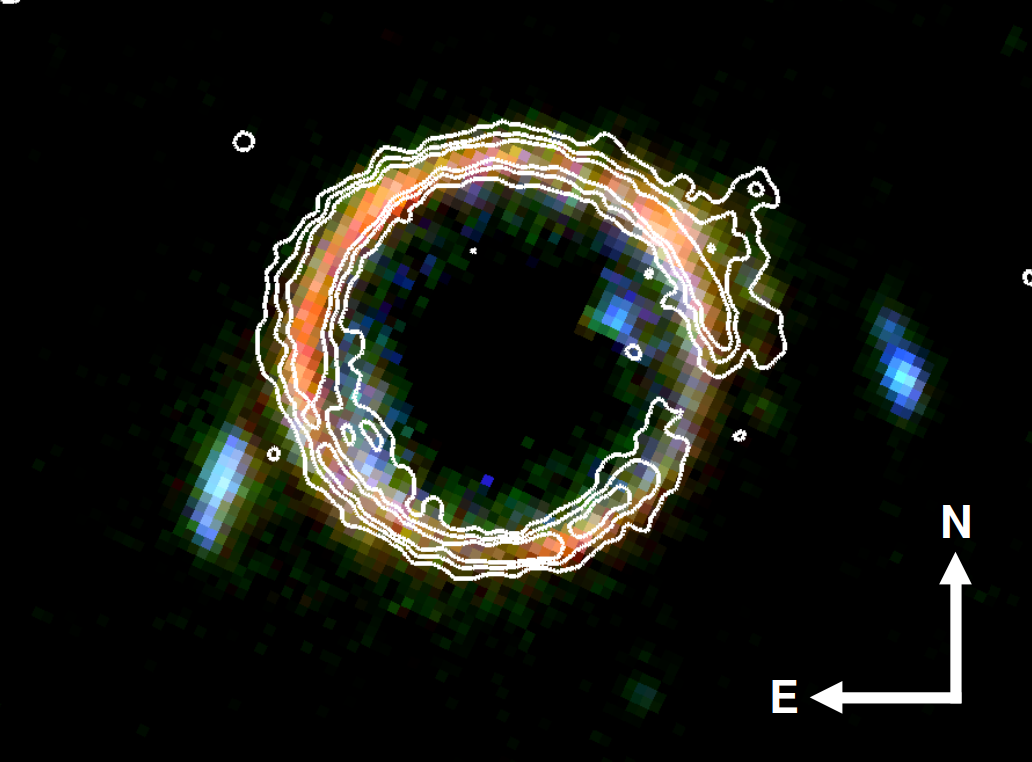}
    \caption{Residual color image of \target\ generated from the F277W, F356W, and F444W \nircam\ imaging after modeling and removal of the foreground lens using GALFIT. 
    The left panel shows \main\ and \companion, with arrows denoting the two images of \companion. \cii contours at the rest frame of \target\ are overlaid on the \nircam\ image in the center panel integrated over the channels where the companion appears. 
    Slight offsets between the ring and the contours in the center panel are due to the contours being integrated only over the channels where the companion appears most strongly, rather than the whole ring. In the right panel we show the \cii contours integrated to show the extended feature opposite the companion, connected to the ring formed by the lensing of \main. The center of the region where the foreground lens was subtracted is masked because the statistical noise is high.}
    \label{fig:galfit}
\end{figure*}

\subsection{\alma} 

We combine ALMA observations of the \cii 158\,\um line and underlying dust continuum observed in projects 2016.1.01374.S (PI: Hezaveh) and 2016.1.01499.S (PI: Litke). Both projects used a similar correlator setup, centering the \cii line in the upper sideband, while the lower sideband provides continuum data. The total on-source time between both projects was 4.9 hours. We performed a continuum subtraction in the $uv$ plane assuming a linear frequency dependence of the continuum emission, excluding frequencies with significant \cii emission from the fit. 

Individual observing blocks from each project span a wide range in spatial resolution from 0.02$^{\prime\prime}$ to 0.15$^{\prime\prime}$. We jointly imaged the data (accounting for the different phase centers and frequency setups of the two projects) using \texttt{tclean} in the  CASA software package \citep{mcmullin2007ASPC..376..127M,casa2022}, combining all line-free frequency channels to create a continuum image. We also imaged the \cii line emission using the continuum-subtracted data with a spectral resolution of 50\,\kms. We base our analysis on images created applying a 50\,mas external taper in the $uv$-plane, which offers a reasonable compromise between spatial resolution and sensitivity and is well-matched to the resolution of the NIRCam imaging. The final synthesized beam size was 93\,mas $\times$ 97\,mas, reaching a sensitivity of 0.20\,mJy/beam in 50\,\kms channels of the \cii cube after correcting for the primary beam response at the position of the lens. We produced integrated and flux-weighted mean velocity maps of the \cii emission after masking pixels detected at $<3\sigma$ significance. 

\begin{figure*}[t]
    \centering
    \includegraphics[width=18cm]{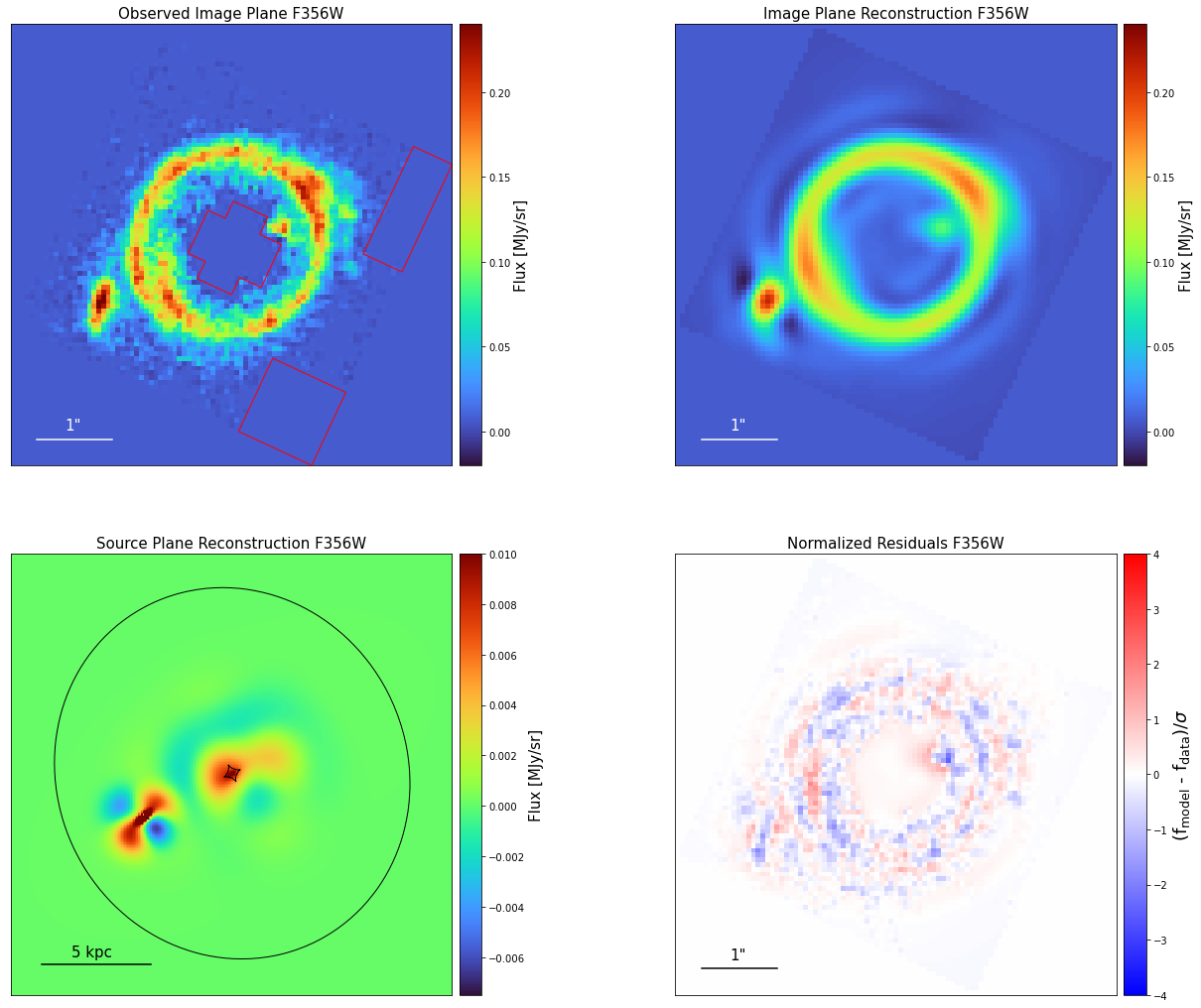}
    \caption{Source plane reconstruction of the F356W band. This band is chosen as the flux from the substructure is strongest relative to the peak of the cores for \main\ and \companion. The  top left panel shows the GALFIT subtracted observed image plane of \main, with regions masked out due to unassociated flux outlined in red. The top right panel is the reconstructed image plane using a combination of parametric models and shapelets.  The bottom left image is a zoomed-in view of the source plane reconstruction with the caustic and critical curve overlayed, while the bottom right panel shows the normalized residuals in the image plane. All panels are rotated 
    such that North is up and East is left. } 
    \label{fig:shapelet}
\end{figure*}

\section{Detection of the Multiply-Imaged Companion}\label{sec:companion}
There are multiple sources near \main\ that could be physically associated if they lie at the same redshift. In this section we investigate each of these sources, then perform a source plane reconstruction of \target.

\subsection{The Multiply-Imaged Companion}

The \jwst\ imaging reveals in detail the stellar emission from \main. 
To better view \target, we  model and remove the flux from the foreground lens, which is an elliptical galaxy at $z=0.263$.
We use GALFIT \citep{peng2002AJ....124..266P,peng2010AJ....139.2097P} to fit a S{\'e}rsic bulge to the foreground lens at the wavelength of the F150W bandpass, convolving the model with the PSF from WebbPSF. In the model we also add a secondary exponential disk component to improve the model in the central region of the galaxy.
During the fitting, we mask nearby galaxies and use the error map to weight the fit. The best fit $\chi^2$ model for the bulge has $n=3.17\pm0.02$ and $r_e=0.88\pm0.01^{\prime\prime}$, while the disk component has $r_e=0.067\pm0.001^{\prime\prime}$ and contains 7\% of the total light.

We take the best fit in F150W, fix the axis ratios, position angles of the two components, and S\'ersic parameter, and then model the lensing galaxy at the other wavelengths, using the appropriate PSFs from WebbPSF. We allow the position, physical scale, and flux to be free parameters. The position is left free to account for 
any remaining subpixel astrometric offsets between images, while the physical scale is left free to account for radial color gradients. 

In the left panel of Figure \ref{fig:galfit}, we show a 3-color image of the lensed galaxy after removal of the foreground lens.\footnote{We mask the noisy residual at the center of the foreground galaxy in this image.} In this image we can now see two bluer objects, one interior to the arc and the other exterior on the opposite side. This geometry is consistent with multiple lensed images of a single companion (\companion), as  
discussed by \citet{peng2022arXiv221016968P}. Proximity of the \main\ and \companion\ in the image plane is suggestive of a small separation in the source plane.  

The \alma\ \cii data also shows clear emission signatures at the locations of the two images of \companion. In the center panel of Figure \ref{fig:galfit} we present contours from the \alma\ \cii data, integrated from $-100$ to $0$ km s$^{-1}$ relative to the central velocity of \main, overlaid on the NIRCam imaging. This emission confirms that  \companion\ is very close to \main\ not only in the image plane, but also in velocity space. Both images of \companion\ have consistent mean relative velocities of approximately $-56$ km s$^{-1}$.
 
\subsection{Other Potential Companions}

In addition to \companion, there are two more  objects visible only in the NIRCam filters at small angular separation. These objects, located at (04:18:39.4,$-$47:51:52.81) and (04:18:39.6,$-$47:51:54.76), 
 are close enough that they would 
both have projected physical separations of $\sim 10$ kpc from \main\ if at the same redshift. These objects do not appear in the ALMA \cii data, 
and photometric redshift estimates using \cigale\ give redshifts of $z_{\text{phot}} = 0.4 \pm 0.25$ and $z_{\text{phot}} < 0.05$ for the West and South objects respectively. We thus conclude that these objects lie in the foreground and are not associated with \main\ or \companion. 

\section{Evidence for an Ongoing Merger}\label{sec:merger}

Given the projected proximity of \companion, we next explore whether this system is an ongoing merger, and attempt to constrain its physical parameters. 

\subsection{Source Plane Reconstruction and Kinematics}

We perform source plane reconstruction using \textsc{lenstronomy} \citep{birrer2018PDU....22..189B,Birrer_2021}. \textsc{lenstronomy} is capable of joint parametric fitting of the foreground lens mass distribution and flux distribution of the background source, and enables simultaneous  fitting of multiple photometric bands. Further, it can use  ``shapelets" \citep{Birrer_2015,Refregier_2003,Refregier_2003b,Massey_2005}, which are a series of 2D 
basis functions, to rapidly reconstruct the source plane without parametric models of the source light, so long as an accurate lens model is used. As a non-parametric method of source plane reconstruction, shapelets can reconstruct substructures that are missed by simple parametric models. Finally, \textsc{lenstronomy} can add shapelets on top of parametric models of the source to find any additional substructure in the source plane without using many orders of shapelets 
for 
structures that are well described by parametric models.
\footnote{One feature of shapelets is that the final models can include regions with unphysical negative flux values. The presence of negative values is intrinsic to the shapelet approach. In the idealized case the summation of the shapelets should result in a net positive flux at all locations, but when modeling with a limited number of shapelets these unphysical negative flux occur at some locations surrounding bright, positive features.}

\begin{figure}[t]
    \centering
    \includegraphics[width=0.48\textwidth]{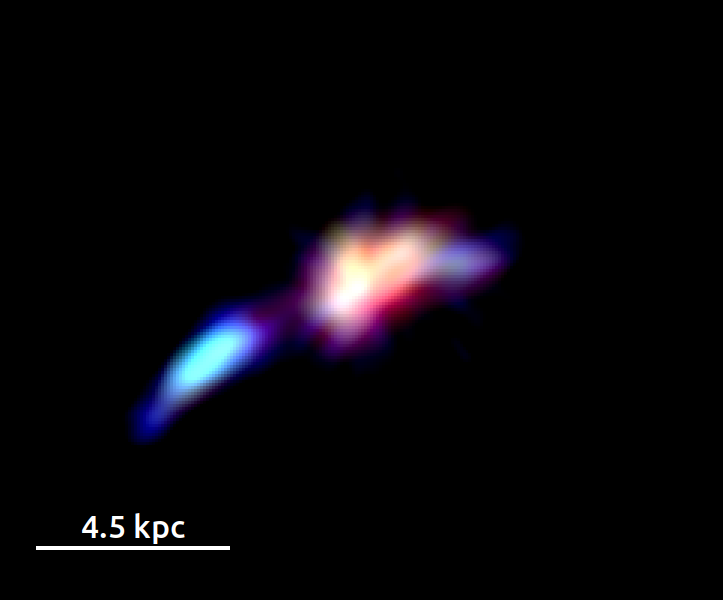}
    \caption{ 
   We show a source plane reconstruction generated using backward ray tracing. We produce this reconstruction using the  F277W, F356W, and F444W filters as in Figure \ref{fig:galfit}. The dust-obscured galaxy \main\ is on the right, and the bluer \companion\ is on the left.}
    \label{fig:raytracing}
\end{figure}

\begin{figure}[t]
    \centering
    \includegraphics[width=0.5\textwidth]{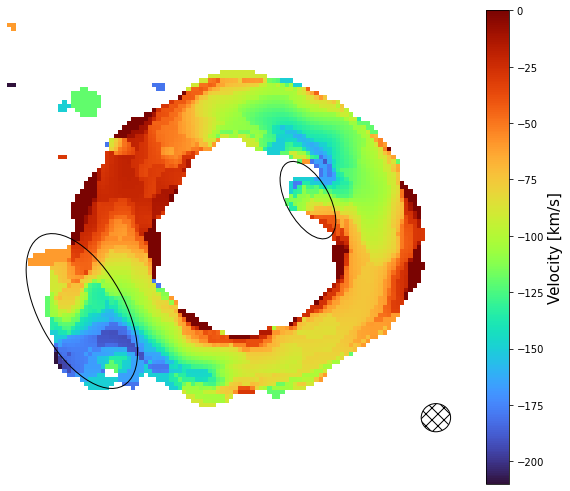}
    \caption{ALMA velocity map of the \target\ system, scaled to highlight the internal velocity gradient of \companion. The two ellipses are the regions where 
    emission from \companion\ is expected, and the area within these two ellipses show mirrored velocity gradients 
    as expected from the lensing geometry. In the bottom right we show the beam size as the cross-hatched ellipse.}
    \label{fig:vel}
\end{figure}

To model the global structure of the \target\ system, we start by using elliptical S{\'e}rsic profiles. 
These simple models provide a good first order fit to the profiles.
To model the mass profile of the lensing galaxy, we use a Singular Isothermal Ellipsoid (SIE) with external shear.  
We perform multi-band fitting of our GALFIT-subtracted data in F277W, F356W, F444W, F560W, and F770W. Starting with  only the parametric profiles, we first pre-sample the space using a Particle Swarm Optimization \citep[PSO][]{Kennedy95} to get near the solution while sampling the high-dimensional space quickly. We then use this approximate solution to seed the Markov Chain Monte Carlo (MCMC) sampler  {\tt emcee} \citep{foreman-mackey2013PASP..125..306F}, allowing it to run to convergence.
We then add shapelets to the source model at each wavelength to better model any substructure for each image. 

In this lens modeling, due to the prior GALFIT subtraction of the lens light, we found difficulty in modeling both \main\ and \companion\ when including the background noise with \textsc{lenstronomy}. In the future, it would be prudent to perform the lens light subtraction within the software used to conduct the lens modeling rather than using two seperate methods. As part of this modeling, we imposed a prior that all values below a threshold of $0.01 \sigma$ less than the median are ignored. This prior forces \textsc{lenstronomy} to consider only flux from regions with higher SNR.

From the source plane reconstruction, we find that the projected physical separation between the centers of the two galaxies is $4.42\pm 0.05$ kpc.
In modeling the source plane with pure S{\'e}rsic profiles, we are unable to reproduce some of the features in both \main\ and \companion. By adding shapelets on top of these S{\'e}rsic profiles, we are able to reproduce those features in the image plane, leading to the reconstruction shown in Figure \ref{fig:shapelet}. 

Figure \ref{fig:shapelet} shows the lens model in the F356W filter, a band in which both  \companion\ and extended structure associated with \main\ can be clearly seen.
We list the parameters for the lens model of this system in Table \ref{tab:lensmodel} along with estimated uncertainties.

Next, we use the parametric   
lens mass model listed in Table \ref{tab:lensmodel} to generate a backwards ray tracing of the system. This approach provides an alternative to the joint parametric - shapelet fitting shown in Figure \ref{fig:shapelet} for assessing the presence of additional structures. 
We take the NIRCAM F277W, F356W, and F444W imaging 
and map it back to the source plane, accounting for magnifications of the multiple images to create the source plane reconstruction. 
The result 
is shown in Figure \ref{fig:raytracing}.  In this Figure, we again see some extended structure in the light distribution, consistent with that inferred from the shapelets. 

Finally, we create a velocity moment map from the \cii data in Figure \ref{fig:vel}. Internal gradients are visible in both \main\ and \companion. 
As highlighted by the ellipses in Figure \ref{fig:vel}, the velocity gradients for the two images of  \companion\
are equal and mirrored, as must be true if we are observing multiple images of the galaxy.

\begin{deluxetable}{cr}
\tablewidth{0pt}
\tablecaption{Lens Model Parameters \label{tab:lensmodel}}
\tablehead{
\colhead{\hspace{1cm}Parameter} & \colhead{\hspace{1cm}Model Value\tablenotemark{a}}
} 
\startdata 
\hspace{1cm}$\theta_E$ &\hspace{1cm} $1.207\pm0.002$\\
\hspace{1cm}$\epsilon_1$ & \hspace{1cm}$-0.031\pm0.006$\\
\hspace{1cm}$\epsilon_2$ & \hspace{1cm}$0.061\pm0.009$\\
\hspace{1cm}$x_\mathrm{Lens}$ & \hspace{1cm}$-0.010\pm0.003$\\ 
\hspace{1cm}$y_\mathrm{Lens}$ & \hspace{1cm}$-0.003\pm0.002$\\
\hspace{1cm}$\gamma_1$ & \hspace{1cm}$-0.005\pm0.003$\\
\hspace{1cm}$\gamma_2$ &\hspace{1cm} $-0.007\pm0.004$\\
\enddata
\tablenotetext{a}{Uncertainties are estimated taking the average of the 16th and 84th percentiles from the marginalized distributions of the MCMC. \textsc{lenstronomy} parameters are defined as follows: $\theta_E$ is the circularized Eistein Radius, $\epsilon_1$ and $\epsilon_2$ are the x and y components to the ellipticity, $x_\mathrm{Lens}$ and $y_\mathrm{Lens}$ are the offset from the center of the image in arcseconds, and $\gamma_1$ and $\gamma_2$ are the external shear components. The orientation of this model is in the native imaging plane, with rotations introduced afterwards as to not introduce autocorrelation noise. The rotation is 245 degrees clockwise from North up and East left.}
\end{deluxetable}

\subsection{Stellar Mass Ratios}
\label{sec:stellarmass}
We next compare the stellar masses of the two components. 
We use the lens model from above (Table \ref{tab:lensmodel}) to derive 
the 
magnifications of \main\ and  \companion. 
We find total magnifications of $\mu=29\pm1$
for \main\ and $\mu=4.2\pm0.9$ for \companion.
The uncertainties on the magnifications are derived from 100 random draws of the MCMC, calculating the flux in the image and the source plane reconstructions of the parametric model. The multiple images of  \companion\ have effective magnifications of $\mu=1.2\pm 0.3$ and $\mu=2.9\pm 0.6$ for the inner and outer image, respectively. 
For \companion\ we use only the data for the source exterior to the \main\ ring to minimize photometric uncertainties due to overlap with the ring.

\begin{figure}[t]
    \centering \includegraphics[width=0.5\textwidth]{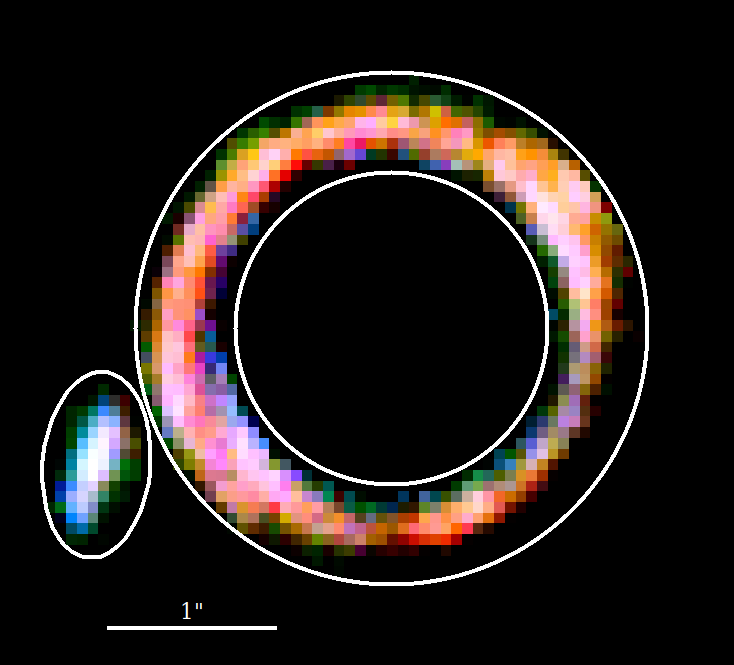}
    \caption{Apertures used for \main\ and  \companion\ galaxy. 
    The inner image of  \companion\ was not used to extract photometry because it is more sensitive to residuals from the subtraction of the lensing galaxy than the outer image.}
    \label{fig:apertures}
\end{figure}

To compare the stellar masses of \main\ and \companion, we extract the photometry for each using fixed photometric apertures. Figure \ref{fig:apertures} shows the apertures used for each component. These apertures are designed to be large enough that there is minimal differential flux loss due to the wavelength dependence of the PSF. 
The flux densities derived for each source are presented in the appendix in Table \ref{tab:imaging}. 
\citet{reuter2020ApJ...902...78R} also provides 100 $\mu$m $-$ 3 mm SPT, LABOCA, Herschel/SPIRE, \alma, and Herschel/PACS data for \target. These flux densities include contributions from both \main\ and \companion. From the ALMA data, we measure that $96\pm 2$\% of the flux is associated with \main. 
When fitting the SED for \main, we include this fraction of the flux density for all bands measured by \citet{reuter2020ApJ...902...78R}. 
The differences in SFR and stellar mass between using this fractional flux and the total flux are smaller than the statistical uncertainties.

\begin{deluxetable*}{cccccccc}[t]
\tablewidth{0pt}
\tablecaption{SED Modeling Results \label{tab:sed}}
\tablehead{
\colhead{Source} & \colhead{Magnification} &\colhead{Code} &  \colhead{Stellar Mass} & \colhead{SFR$_{100}$} & \colhead{Z$_*$ } & \colhead{A$_V$ } &\colhead{sSFR}\\[-0.25cm]
 \colhead{} & \colhead{$\mu$} & \colhead{} &\colhead{($\mu$M$_{\odot}$)} & \colhead{($\mu$M$_{\odot}$/yr)} & \colhead{ (Z$_{\odot}$)} & \colhead{} & \colhead{($10^{-9}$~M$_{\odot}~\mathrm{yr}^{-1}$)}
}
\startdata 
\main\ & $29.5\pm1.2$ & \prospector\ & 
10.2$^{+3.4}_{-1.9}\times10^{11}$ & 4257$^{+989}_{-742}$ & 0.94$^{+0.15}_{-0.11}$ & 3.75$^{+0.12}_{-0.14}$ & $4.2^{+1.0}_{-0.7}$\\
  &  & \cigale\ & 4.5$\pm$0.9$\times10^{11}$ & 3770$\pm$545 & 1 & 3.8$\pm$0.1 & $8.4\pm1.2$\\
\companion\ & $2.92\pm0.63$ & \prospector\  & 
23.9$^{+9.1}_{-7.2}\times10^{9}$ & 25.3$^{+23.6}_{-15.2}$ & 0.68$^{+0.04}_{-0.02}$ & 1.5$^{+0.32}_{-0.43}$& $1.1^{+1.1}_{-0.7}$\\
& & \cigale\ & 9.8$\pm0.5\times10^{9}$ & 43.6$\pm$17.9 & 1 & 1.4$\pm$0.2&$4.4\pm1.8$\\
\enddata
\tablecomments{Quoted parameter results are values without any magnification correction. The 
\cigale\ models assume a fixed stellar metallicity value set to $Z_{\odot}$, as this code only permits a few discrete values. 
}
\end{deluxetable*}

We use both \cigale\  \citep{boquien2019A&A...622A.103B} and \prospector\  \citep{johnson2021ApJS..254...22J} to fit the SEDs. This provides a measure of systematic uncertainties associated with assumptions in the SED modeling.
For both codes we assume a Chabrier IMF \citep{chabrier2003PASP..115..763C}. For both codes we use a flexible attenuation curve parameterization that allows for a variable UV-optical slope and $V$-band attenuation: Specifically, \cigale\ uses the parameterization from \citet{boquien2019A&A...622A.103B} and \prospector\ that of \citet{kriek2013ApJ...775L..16K}.
The metallicity is allowed to vary for \main, and we apply a prior that the metallicity must be between 80\% and 125\% of solar, 
consistent with \citet{debreuck2019A&A...631A.167D}, \citet{peng2022arXiv221016968P}, and our team’s in-progress spectroscopic analysis (Birkin et al., submitted).

For \companion, we set a uniform prior between 65$-$75\% solar on the metallicity for \prospector\  based upon the NIRSpec analysis of Birkin et al. (in prep). We fix the metallicity to solar for \cigale, as this is the closest metallicity value available in the code.
With \cigale, we model the SFH parametrically as an exponential decay ($\tau_{\text{best}}$ = 1 Gyr) with an additional burst component using the \cite{bruzual_charlot_sps} stellar population model.
In contrast, \prospector\   uses  Flexible Stellar Population Synthesis (FSPS) stellar population modeling \citep{fsps_1, fsps_2} and we allow a non-parametric 
SFH with a Dirichlet prior on the mass formed per time bin \citep{leja2019_prospector_sfhs}. \cite{lower2020ApJ...904...33L} demonstrated with simulations that such an approach can outperform parametric models. The \prospector\  fit also permits a non-uniform dust screen, and we let the fractional obscuration be a free parameter \citep{lower_2022_dust_attn}.  The posterior for the obscured fraction peaks at $>98\%$ for \main; however, the small amount of light that is unobscured is important in the SED fit.

\begin{figure*}[t]
    \centering
    \includegraphics[width=0.48\textwidth]{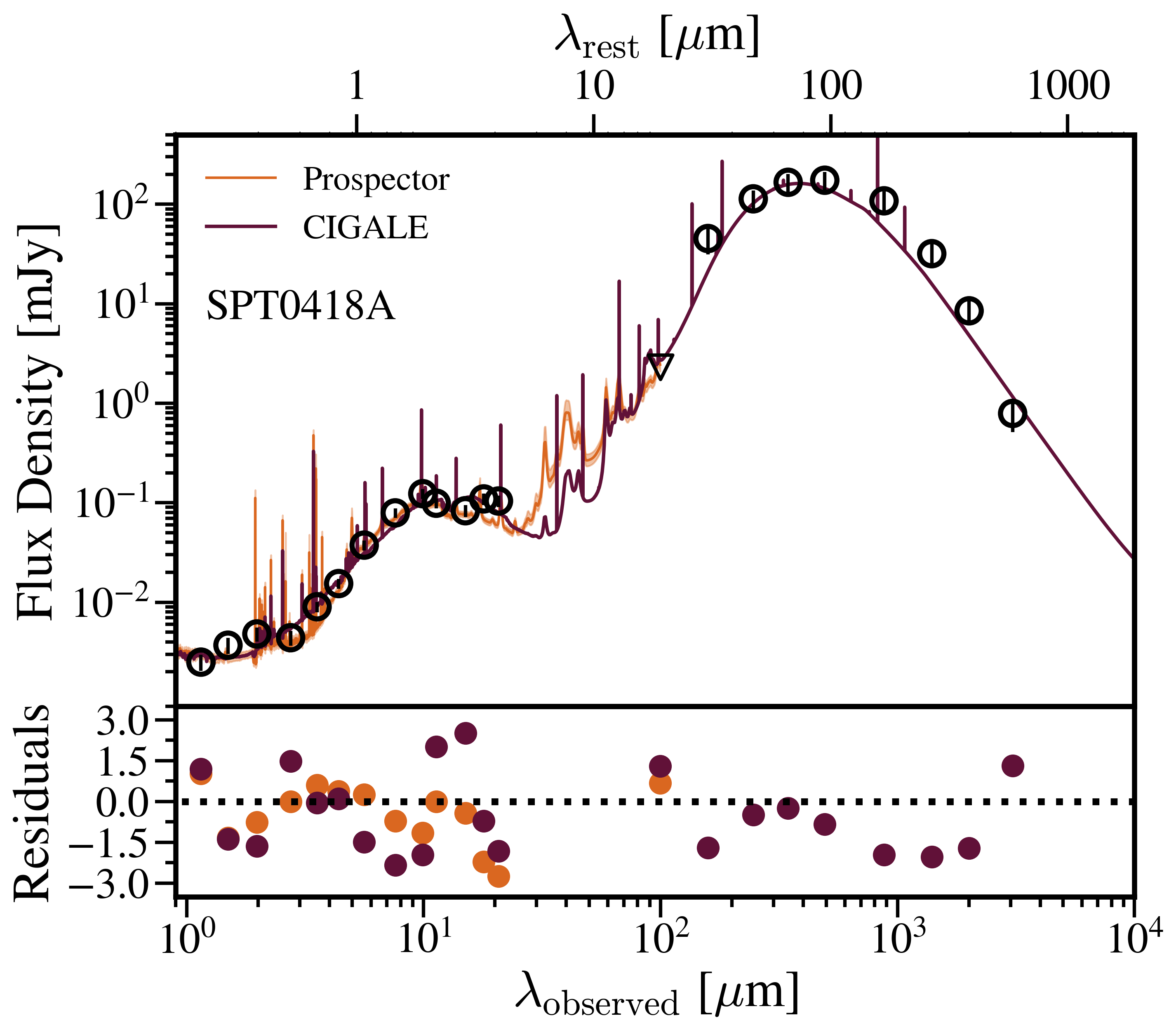}
    \includegraphics[width=0.48\textwidth]{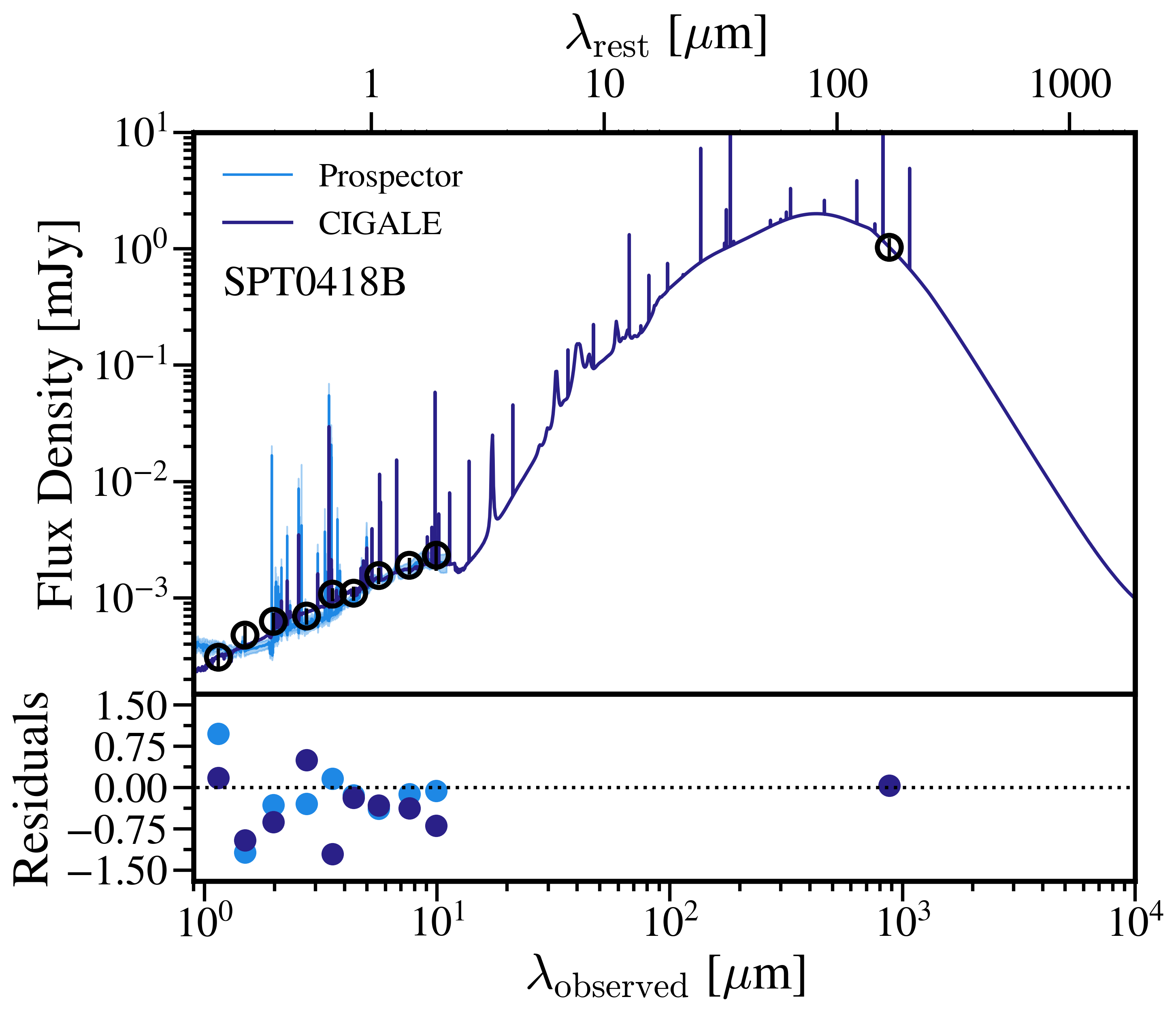}\\
    \includegraphics[width=0.48\textwidth]{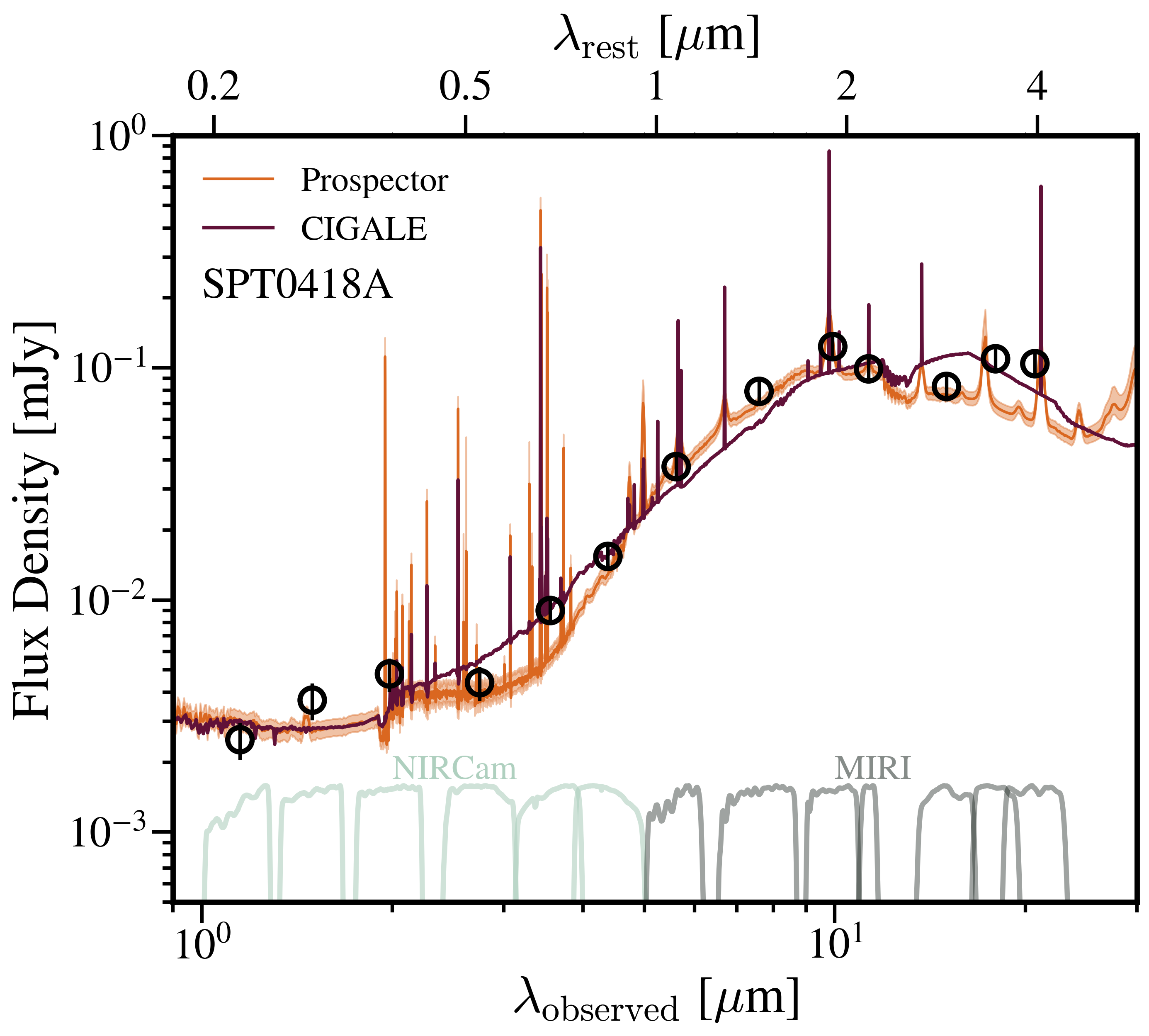}
    \includegraphics[width=0.48\textwidth]{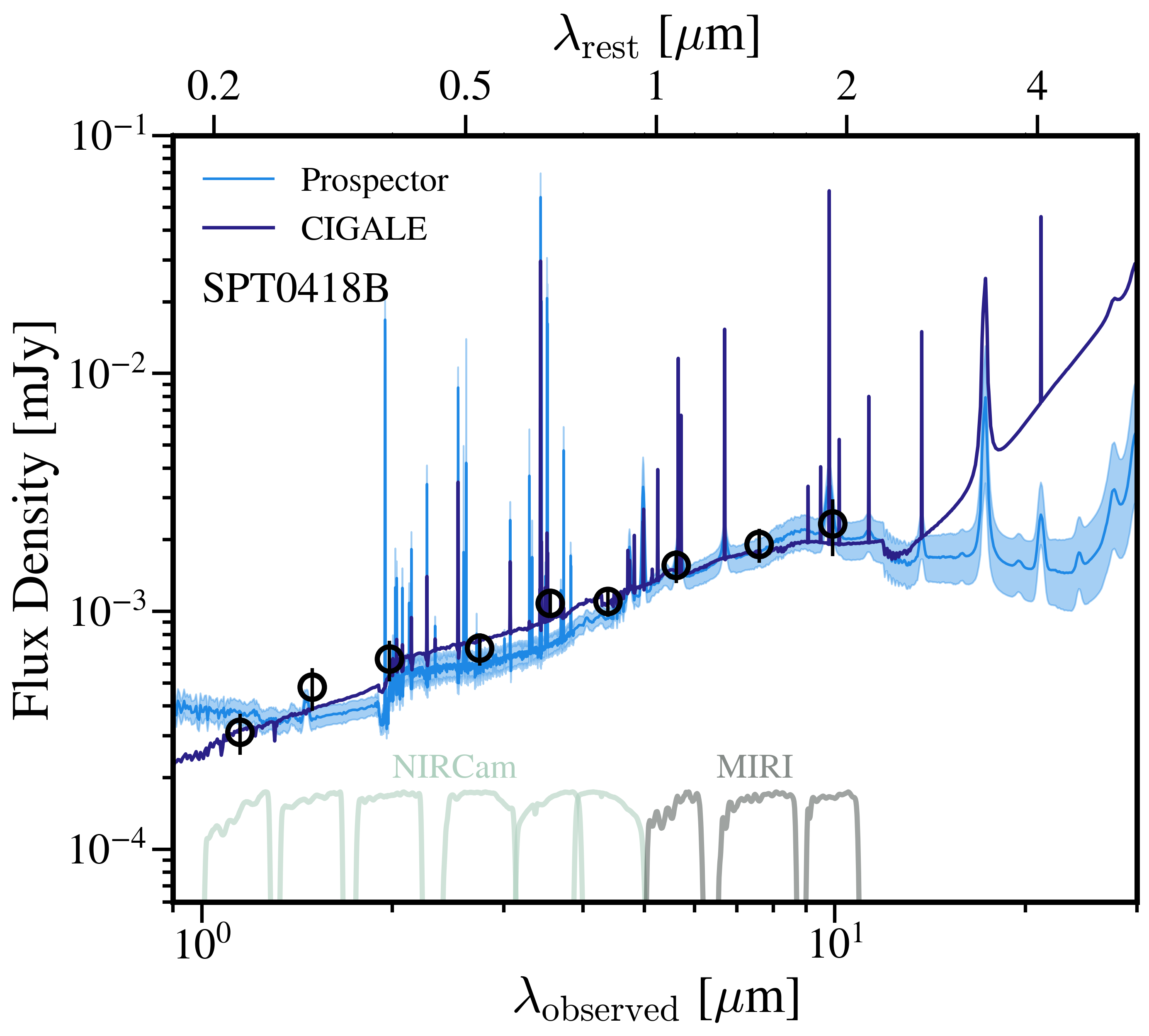}
    \caption{Top: UV-FIR Spectral energy distributions (SEDs) for \main\ (left) and \companion\ (right), with observed photometry not corrected for magnification. The best-fit models from \prospector\  and \cigale\  are shown as shaded bands and solid curves, respectively. 
    The \prospector\ SED is 
    the median model SED with the $16^{th} - 84^{th}$ percentiles shown in the shaded regions, while the best-fit \cigale\  SED is the model corresponding to the minimum $\chi^2$. The corresponding model parameters can be found in Table \ref{tab:sed}. Bottom: Zoom-in to rest-frame UV and optical.}
    \label{fig:sed}
\end{figure*}

In Figure \ref{fig:sed} we show the SED models that best fit our data with each code for both \main\ and  \companion. Similarly, the derived quantities from these SED Models are presented in Table \ref{tab:sed}. The constraints on \companion\ are less robust due to the more limited set of photometric detections since this source is relatively faint and the \miri\ integration times are short. For \prospector, we plot the median and $1\sigma$ spread in the model SEDs, and for \cigale\ we plot the model corresponding to the minimum $\chi^2$.\footnote{\cigale\ does not calculate the $1\sigma$ spread in model SEDs.} For \prospector, we find that it fits the FIR dust emission poorly. Thus, we had \prospector\ model only the stellar emission of the SED as to not have the improper dust models bias the fit. The reduced $\chi^2$ statistics for the \cigale\ fits to the \main\ and \companion\ SEDs are $3.35$ and $0.48$, respectively, and  $3.42$ and $0.54$ when fit with \prospector. The differences in $\chi^2$ values between the two sources are caused by both number of observations and models. Since \companion\ only has one data point in the FIR, it overfits the point for all of the FIR emission. \footnote{We note that the $\chi^2$ values for the two SED fitting packages are not directly comparable because the codes use different methodologies to find the best fit SEDs.}

A comparison of the results from the two SED modeling codes in Table \ref{tab:sed} shows that the derived star formation rates and effective $A_V$ from \cigale\ and \prospector\  are consistent within the uncertainties for both \main\ and \companion.
The inferred stellar metallicity for the main source is consistent with the solar value we have assumed in the \cigale\ model.

The most significant discrepancy between the two codes is with the inferred stellar mass. For \main, \cigale\  infers a magnification-corrected stellar mass of 1.5$\pm0.6\times10^{10}$ M$_\odot$,  which is a factor of 2.3 times smaller than the \prospector\  median stellar mass value of 3.5$^{+1.4}_{-1.4}\times10^{10}$ M$_\odot$. This increases to a factor of 3.9 smaller for  \companion, for which \cigale\ yields 3.4$\pm0.7\times10^{9}$ M$_\odot$ compared to 8.2$^{+3.6}_{-3.0}\times10^{9}$ M$_\odot$ for \prospector. These differences in the inferred stellar masses are largely driven by the different models for the star formation history: in \cigale, we assumed a declining exponential SFH with a burst component, with an inferred $80$\% of the galaxy's stellar mass formed in the recent burst. The non-parametric SFH model we used in \prospector\ does not \textit{a priori} assume a shape for the SFH, rather it fits for the stellar mass formed in each bin and estimates that only $40\%$ of the total galaxy stellar mass was formed in a recent burst, with a significant fraction of the stellar mass having formed at earlier times. We verified that if we require the SFH to be the same in both codes, then the two codes give consistent results.

Setting aside the offset between the two codes for the derived stellar masses, the 
ratio of the stellar masses for \main\ and \companion\ is consistent between the two codes. 
Accounting for magnification, the inferred true stellar mass ratios are $4.5\pm 1.0$ for \cigale\  and $4.2^{+1.9}_{-1.6}$ for \prospector, respectively.
The uncertainties are large, but still enable us to
conclude that this system is consistent with being a $\sim 4$ to 1 ongoing minor merger between the two galaxies. In this context, the extended structure associated with \main\ is potentially  a tidal feature arising from the interaction.

It is interesting to consider the effect of this merger upon the stellar populations in these galaxies. \citet{rizzo2020Natur.584..201R} found a dynamically well-ordered disk in \main, which would argue that the interaction has not significantly influenced the internal dynamics. The specific star formation rate (sSFR) of \main\ is consistent with estimates for the star-forming main sequence at $z=4$ as found in \citet{bouwens2012ApJ...754...83B} for UV dropout Lyman break galaxies, and similar to what is seen by \citet{dacunha2015ApJ...806..110D} for ALESS galaxies. The sSFR of \main\ therefore appears to not be significantly elevated by the merger.  The specific SFR (sSFR) of  \companion\ is also $<10^{-8}$ yr$^{-1}$ from both codes, comparable to typical values found for star-forming galaxies at this redshift \citep[see][and references therein]{heinis2014MNRAS.437.1268H}. Interestingly, despite this ongoing collision, there is no evidence that the merger event is actively elevating the star formation in either galaxy.

\section{Conclusions}\label{sec:conclusion}

In this paper we use \jwst\ \nircam\ and \miri\ imaging of \target, which is at a redshift ($z=4.225$), from the TEMPLATES Early Release Science program to investigate the unlensed properties of the \target\ system. 
We analyze the stellar emission from both \main\ and the multiply-imaged companion galaxy (\companion), deriving a lensing model and determining stellar masses. \cii data from ALMA confirms that \companion\ lies at the same redshift as \main, which was also reported in \citet{peng2022arXiv221016968P}, and exhibits a velocity gradient consistent with internal rotation. 

From our lensing model reconstruction of the source plane, we determine that the projected physical offset between the centroids of the \main\ and \companion. We measure a value of $4.42\pm 0.05$ kpc in the source plane, which is broadly consistent with the estimate of \citet{peng2022arXiv221016968P}. We also see tentatively evidence for extended features associated with \main.
The ALMA data also demonstrate that the mean velocity offset between the two galaxies is $-56$ km s$^{-1}$. 
Fits to the SEDs of both sources using both \prospector\ and \cigale\  indicate that the stellar mass ratio of the two galaxies is approximately 4 to 1 --- $4.5\pm 1.0$ for \cigale\  and $4.2^{+1.9}_{-1.6}$ for \prospector. 
The simplest interpretation of these results is that 
we are witnessing an ongoing minor merger in this system. Despite this merger, there is no evidence for elevated sSFRs in either galaxy. Thus, the two galaxies are consistent with the star-formation main sequence at their redshift. 
After applying the magnification correction, we find that \main\ 
has a stellar mass of  $M_\star=3.4^{+1.4}_{-1.4}\times10^{10}$ M$_\odot$ from \prospector, or $M_\star=1.5\pm0.6\times10^{10}$ M$_\odot$ from \cigale, and \companion\ has a stellar mass of $M_\star=8.2^{+3.6}_{-3.0}\times10^{9}$ M$_\odot$ from \prospector, or $M_\star=3.4\pm0.7\times10^{9}$ M$_\odot$ from \cigale. We suggest that the difference between these two codes is due to different star assumed formation histories found by each.

Finally, we have compared the results of fitting the SEDs with \prospector\ and \cigale. The two  codes yield similar SFRs, effective extinctions (Table \ref{tab:sed}), and stellar mass ratios.  \prospector\  however yields stellar masses that are systematically higher by factors of  $\sim2-4$. This difference, which is largely driven by differences in the models of the star formation history, highlights that caution should be exercised when comparing
stellar mass estimates in the literature that use different codes and input assumptions \citep[see also][]{michalowski2014A&A...571A..75M,mobasher2015ApJ...808..101M,hunt2019A&A...621A..51H}.

\section{Acknowledgments}

 This work is based in part on observations made with the NASA/ESA/CSA JWST. The data were obtained from the Mikulski Archive for Space Telescopes at the Space Telescope Science Institute, which is operated by the Association of Universities for Research in Astronomy, Inc., under NASA contract NAS 5-03127 for \jwst. These observations are associated with ERS program \#1355. We express our gratitude to the thousands of people around the world who brought to fruition \jwst\ and its science instruments \nircam\ and \miri.

 The SPT is supported by the NSF through grant OPP-1852617.
 This paper makes use of the following ALMA data: ADS/NSF.ALMA\#2016.1.01374.S and \#2016.1.01499.S  ALMA is a partnership of ESO (representing its member states), NSF (USA) and NINS (Japan), together with NRC (Canada), MOST and ASIAA (Taiwan), and KASI (Republic of Korea), in cooperation with the Republic of Chile. The Joint ALMA Observatory is operated by ESO, AUI/NRAO and NAOJ. The National Radio Astronomy Observatory is a facility of the National Science Foundation operated under cooperative agreement by Associated Universities, Inc.
 
 Support for this program was provided by NASA through a grant from the Space Telescope Science Institute (JWST-ERS-01355), which is operated by the Association of Universities for Research in Astronomy, Inc., under NASA contract NAS 5-03127. M.A. acknowledges support from FONDECYT grant 1211951 and CONICYT + PCI + INSTITUTO MAX PLANCK DE ASTRONOMIA MPG190030. 
 D.P.M. J.D.V., and K.P. acknowledge support from the US NSF under grants AST-1715213 and AST-1716127.
 MA and MS acknowledge support from and CONICYT + PCI + REDES 190194, and ANID BASAL project FB210003. D.N. was supported by the NSF via AST-1909153. K.A.P. is supported by the Center for AstroPhysical Surveys at the National Center for Supercomputing Applications as an Illinois Survey Science Graduate Fellow. N.S. is a member of the International Max Planck Research School (IMPRS) for Astronomy and Astrophysics at the Universities of Bonn and Cologne. 

\appendix

\section{Photometry}
We present the photometric data used in modeling the SEDs in Figure \ref{fig:sed}. Table \ref{tab:imaging} includes the \jwst\ photometry from NIRCam and MIRI, while Table \ref{tab:FIR} summarizes the longer wavelength literature fluxes from \citet{reuter2020ApJ...902...78R}. The values in Table \ref{tab:FIR} correspond to the total flux from \main\ and \companion. As discussed in the text, when fitting the SEDs for the two components we assign $95.8\pm 2.2$\% of the flux in the FIR bands to \main. This percentage corresponds to the fraction of the flux associated with \main\ in the ALMA imaging, and we assume the same fractional contribution for the other FIR observations.

\begin{deluxetable}{ccccrr}[h]
\tablewidth{0pt} 
\tablecaption{NIRCam and MIRI Data \label{tab:imaging}}
\tablehead{
\colhead{Filter} & \colhead{$\lambda$\tablenotemark{a}} & \colhead{$\Delta\lambda$} & \colhead{Int.~Time} & \colhead{SPT 0418A} & \colhead{SPT 0418B}\\[-0.25cm]
\colhead{} & \colhead{$\mu$m}&\colhead{$\mu$m} &\colhead{(s)} & \colhead{($\mu$Jy)} & \colhead{ ($\mu$Jy)}
} 
\startdata 
F115W &1.154 &0.225& 687 & 2.5$\pm$0.2 & 0.31$\pm$0.03\\
F150W &1.501&0.318& 343 & 3.7$\pm$0.3 & 0.48$\pm$0.05 \\
F200W &1.990&0.461& 429 & 4.8$\pm$0.3 & 0.63$\pm$0.06\\
F277W &2.786&0.672& 687 & 4.4$\pm$0.3 & 0.70$\pm$0.04\\ 
F356W &3.563&0.787& 343 & 9.0$\pm$0.2& 1.08$\pm$0.03\\
F444W &4.421&1.024& 429 & 15.4$\pm$0.2 & 1.10$\pm$0.04\\
\hline
F560W &5.6&1.2& 277 & 37.5$\pm$0.5 & 1.56$\pm$0.09 \\
F770W &7.7&2.2& 144 & 79.1$\pm$0.9 & 1.91$\pm$0.12\\
F1000W &10&2.0& 111 & 123.3$\pm$2.2 & 2.33$\pm$0.39 \\
F1280W &12.8&2.4& 111 & 98.9$\pm$1.4& \nodata \\
F1500W &15&3.0& 111 & 83.1$\pm$3.3 & \nodata \\
F1800W &18&3.0& 222 & 109.3$\pm$3.6& \nodata \\
F2100W &22&5.0& 832 & 104.1$\pm$3.4& \nodata \\
\enddata
\tablenotetext{a}{For NIRCam $\lambda$ refers to the pivot wavelength, while for \miri\ $\lambda$ is simply the central wavelength.}
\tablecomments{Quoted flux densities are observed values without any magnification correction. The fluxes for the \companion\ are for the outer image alone.  See \S \ref{sec:stellarmass} for more details on the photometry measurements. }
\end{deluxetable}

\begin{deluxetable}{cc}[]
\tablewidth{\textwidth} 
\tablecaption{Published \target\ FIR 
Photometry 
from \citet{reuter2020ApJ...902...78R} \label{tab:FIR}}
\tablehead{
\colhead{\hspace{1cm}$\lambda$} &  \colhead{\hspace{1cm}Flux Density} \\
\colhead{\hspace{1cm}($\mu$m)}  & \colhead{\hspace{1cm}(mJy)\hspace{0.5cm}}
 }
\startdata 
\hspace{1cm}100 & \hspace{1cm}$<7$\hspace{0.5cm}\\
\hspace{1cm}160 & \hspace{1cm}$45\pm8$\hspace{0.5cm}\\
\hspace{1cm}250 &\hspace{1cm}$114\pm6$\hspace{0.5cm}\\
\hspace{1cm}350 &\hspace{1cm}$166\pm6$\hspace{0.5cm}\\ 
\hspace{1cm}500 &\hspace{1cm}$175\pm7$\hspace{0.5cm}\\
\hspace{1cm}870 &\hspace{1cm}$108\pm11$\hspace{0.5cm}\\
\hspace{1cm}1400 & \hspace{1cm}$32\pm5$\hspace{0.5cm}\\
\hspace{1cm}2000 & \hspace{1cm}$9\pm1$\hspace{0.5cm}\\
\hspace{1cm}3000 &\hspace{1cm} $0.79\pm0.14$\hspace{0.5cm}\\
\enddata
\end{deluxetable}

\bibliography{spt0418}{}

\begin{thebibliography}{}
\expandafter\ifx\csname natexlab\endcsname\relax\def\natexlab#1{#1}\fi
\providecommand{\url}[1]{\href{#1}{#1}}
\providecommand{\dodoi}[1]{doi:~\href{http://doi.org/#1}{\nolinkurl{#1}}}
\providecommand{\doeprint}[1]{\href{http://ascl.net/#1}{\nolinkurl{http://ascl.net/#1}}}
\providecommand{\doarXiv}[1]{\href{https://arxiv.org/abs/#1}{\nolinkurl{https://arxiv.org/abs/#1}}}

\bibitem[{{Alaghband-Zadeh} {et~al.}(2012){Alaghband-Zadeh}, {Chapman},
  {Swinbank}, {Smail}, {Harrison}, {Alexander}, {Casey}, {Dav{\'e}},
  {Narayanan}, {Tamura}, \& {Umehata}}]{alaghband2012MNRAS.424.2232A}
{Alaghband-Zadeh}, S., {Chapman}, S.~C., {Swinbank}, A.~M., {et~al.} 2012,
  \mnras, 424, 2232, \dodoi{10.1111/j.1365-2966.2012.21386.x}

\bibitem[{{Baugh} {et~al.}(1996){Baugh}, {Cole}, \&
  {Frenk}}]{baugh1996MNRAS.283.1361B}
{Baugh}, C.~M., {Cole}, S., \& {Frenk}, C.~S. 1996, \mnras, 283, 1361,
  \dodoi{10.1093/mnras/283.4.1361}

\bibitem[{{Birrer} \& {Amara}(2018)}]{birrer2018PDU....22..189B}
{Birrer}, S., \& {Amara}, A. 2018, Physics of the Dark Universe, 22, 189,
  \dodoi{10.1016/j.dark.2018.11.002}

\bibitem[{Birrer {et~al.}(2015)Birrer, Amara, \& Refregier}]{Birrer_2015}
Birrer, S., Amara, A., \& Refregier, A. 2015, The Astrophysical Journal, 813,
  102, \dodoi{10.1088/0004-637x/813/2/102}

\bibitem[{Birrer {et~al.}(2021)Birrer, Shajib, Gilman, Galan, Aalbers, Millon,
  Morgan, Pagano, Park, Teodori, Tessore, Ueland, de~Vyvere, Wagner-Carena,
  Wempe, Yang, Ding, Schmidt, Sluse, Zhang, \& Amara}]{Birrer_2021}
Birrer, S., Shajib, A., Gilman, D., {et~al.} 2021, Journal of Open Source
  Software, 6, 3283, \dodoi{10.21105/joss.03283}

\bibitem[{{Boquien} {et~al.}(2019){Boquien}, {Burgarella}, {Roehlly}, {Buat},
  {Ciesla}, {Corre}, {Inoue}, \& {Salas}}]{boquien2019A&A...622A.103B}
{Boquien}, M., {Burgarella}, D., {Roehlly}, Y., {et~al.} 2019, \aap, 622, A103,
  \dodoi{10.1051/0004-6361/201834156}

\bibitem[{{Bouwens} {et~al.}(2012){Bouwens}, {Illingworth}, {Oesch}, {Franx},
  {Labb{\'e}}, {Trenti}, {van Dokkum}, {Carollo}, {Gonz{\'a}lez}, {Smit}, \&
  {Magee}}]{bouwens2012ApJ...754...83B}
{Bouwens}, R.~J., {Illingworth}, G.~D., {Oesch}, P.~A., {et~al.} 2012, \apj,
  754, 83, \dodoi{10.1088/0004-637X/754/2/83}

\bibitem[{{Boyer} {et~al.}(2022){Boyer}, {Anderson}, {Gennaro}, {Geha},
  {Wingfield McQuinn}, {Tollerud}, {Correnti}, {Brenner Newman}, {Cohen},
  {Kallivayalil}, {Beaton}, {Cole}, {Dolphin}, {Kalirai}, {Sandstrom},
  {Savino}, {Skillman}, {Weisz}, \& {Williams}}]{boyer2022}
{Boyer}, M.~L., {Anderson}, J., {Gennaro}, M., {et~al.} 2022, Research Notes of
  the American Astronomical Society, 6, 191, \dodoi{10.3847/2515-5172/ac923a}

\bibitem[{{Bruzual} \& {Charlot}(2003)}]{bruzual_charlot_sps}
{Bruzual}, G., \& {Charlot}, S. 2003, \mnras, 344, 1000,
  \dodoi{10.1046/j.1365-8711.2003.06897.x}

\bibitem[{{Chabrier}(2003)}]{chabrier2003PASP..115..763C}
{Chabrier}, G. 2003, \pasp, 115, 763, \dodoi{10.1086/376392}

\bibitem[{Conroy \& Gunn(2010)}]{fsps_2}
Conroy, C., \& Gunn, J.~E. 2010, The Astrophysical Journal, 712, 833–857,
  \dodoi{10.1088/0004-637x/712/2/833}

\bibitem[{Conroy {et~al.}(2009)Conroy, Gunn, \& White}]{fsps_1}
Conroy, C., Gunn, J.~E., \& White, M. 2009, The Astrophysical Journal, 699,
  486–506, \dodoi{10.1088/0004-637x/699/1/486}

\bibitem[{{da Cunha} {et~al.}(2015){da Cunha}, {Walter}, {Smail}, {Swinbank},
  {Simpson}, {Decarli}, {Hodge}, {Weiss}, {van der Werf}, {Bertoldi},
  {Chapman}, {Cox}, {Danielson}, {Dannerbauer}, {Greve}, {Ivison}, {Karim}, \&
  {Thomson}}]{dacunha2015ApJ...806..110D}
{da Cunha}, E., {Walter}, F., {Smail}, I.~R., {et~al.} 2015, \apj, 806, 110,
  \dodoi{10.1088/0004-637X/806/1/110}

\bibitem[{{De Breuck} {et~al.}(2019){De Breuck}, {Wei{\ss}}, {B{\'e}thermin},
  {Cunningham}, {Apostolovski}, {Aravena}, {Archipley}, {Chapman}, {Chen},
  {Fu}, {Jarugula}, {Malkan}, {Mangian}, {Phadke}, {Reuter}, {Stacey},
  {Strandet}, {Vieira}, \& {Vishwas}}]{debreuck2019A&A...631A.167D}
{De Breuck}, C., {Wei{\ss}}, A., {B{\'e}thermin}, M., {et~al.} 2019, \aap, 631,
  A167, \dodoi{10.1051/0004-6361/201936169}

\bibitem[{{Engel} {et~al.}(2010){Engel}, {Tacconi}, {Davies}, {Neri}, {Smail},
  {Chapman}, {Genzel}, {Cox}, {Greve}, {Ivison}, {Blain}, {Bertoldi}, \&
  {Omont}}]{engel2010ApJ...724..233E}
{Engel}, H., {Tacconi}, L.~J., {Davies}, R.~I., {et~al.} 2010, \apj, 724, 233,
  \dodoi{10.1088/0004-637X/724/1/233}

\bibitem[{{Foreman-Mackey} {et~al.}(2013){Foreman-Mackey}, {Hogg}, {Lang}, \&
  {Goodman}}]{foreman-mackey2013PASP..125..306F}
{Foreman-Mackey}, D., {Hogg}, D.~W., {Lang}, D., \& {Goodman}, J. 2013, \pasp,
  125, 306, \dodoi{10.1086/670067}

\bibitem[{{Governato} {et~al.}(2009){Governato}, {Brook}, {Brooks}, {Mayer},
  {Willman}, {Jonsson}, {Stilp}, {Pope}, {Christensen}, {Wadsley}, \&
  {Quinn}}]{governato2009MNRAS.398..312G}
{Governato}, F., {Brook}, C.~B., {Brooks}, A.~M., {et~al.} 2009, \mnras, 398,
  312, \dodoi{10.1111/j.1365-2966.2009.15143.x}

\bibitem[{{Heinis} {et~al.}(2014){Heinis}, {Buat}, {B{\'e}thermin}, {Bock},
  {Burgarella}, {Conley}, {Cooray}, {Farrah}, {Ilbert}, {Magdis}, {Marsden},
  {Oliver}, {Rigopoulou}, {Roehlly}, {Schulz}, {Symeonidis}, {Viero}, {Xu}, \&
  {Zemcov}}]{heinis2014MNRAS.437.1268H}
{Heinis}, S., {Buat}, V., {B{\'e}thermin}, M., {et~al.} 2014, \mnras, 437,
  1268, \dodoi{10.1093/mnras/stt1960}

\bibitem[{{Hopkins} {et~al.}(2008){Hopkins}, {Hernquist}, {Cox}, \&
  {Kere{\v{s}}}}]{hopkins2008ApJS..175..356H}
{Hopkins}, P.~F., {Hernquist}, L., {Cox}, T.~J., \& {Kere{\v{s}}}, D. 2008,
  \apjs, 175, 356, \dodoi{10.1086/524362}

\bibitem[{{Hopkins} {et~al.}(2010){Hopkins}, {Bundy}, {Croton}, {Hernquist},
  {Keres}, {Khochfar}, {Stewart}, {Wetzel}, \&
  {Younger}}]{hopkins2010ApJ...715..202H}
{Hopkins}, P.~F., {Bundy}, K., {Croton}, D., {et~al.} 2010, \apj, 715, 202,
  \dodoi{10.1088/0004-637X/715/1/202}

\bibitem[{{Hunt} {et~al.}(2019){Hunt}, {De Looze}, {Boquien}, {Nikutta},
  {Rossi}, {Bianchi}, {Dale}, {Granato}, {Kennicutt}, {Silva}, {Ciesla},
  {Rela{\~n}o}, {Viaene}, {Brandl}, {Calzetti}, {Croxall}, {Draine},
  {Galametz}, {Gordon}, {Groves}, {Helou}, {Herrera-Camus}, {Hinz}, {Koda},
  {Salim}, {Sandstrom}, {Smith}, {Wilson}, \&
  {Zibetti}}]{hunt2019A&A...621A..51H}
{Hunt}, L.~K., {De Looze}, I., {Boquien}, M., {et~al.} 2019, \aap, 621, A51,
  \dodoi{10.1051/0004-6361/201834212}

\bibitem[{{Johnson} {et~al.}(2021){Johnson}, {Leja}, {Conroy}, \&
  {Speagle}}]{johnson2021ApJS..254...22J}
{Johnson}, B.~D., {Leja}, J., {Conroy}, C., \& {Speagle}, J.~S. 2021, \apjs,
  254, 22, \dodoi{10.3847/1538-4365/abef67}

\bibitem[{Kennedy \& Eberhart(1995)}]{Kennedy95}
Kennedy, J., \& Eberhart, R. 1995, in Proceedings of ICNN'95 - International
  Conference on Neural Networks, Vol.~4, 1942--1948 vol.4,
  \dodoi{10.1109/ICNN.1995.488968}

\bibitem[{{Kriek} \& {Conroy}(2013)}]{kriek2013ApJ...775L..16K}
{Kriek}, M., \& {Conroy}, C. 2013, \apjl, 775, L16,
  \dodoi{10.1088/2041-8205/775/1/L16}

\bibitem[{{Leja} {et~al.}(2019){Leja}, {Carnall}, {Johnson}, {Conroy}, \&
  {Speagle}}]{leja2019_prospector_sfhs}
{Leja}, J., {Carnall}, A.~C., {Johnson}, B.~D., {Conroy}, C., \& {Speagle},
  J.~S. 2019, \apj, 876, 3, \dodoi{10.3847/1538-4357/ab133c}

\bibitem[{{Litke} {et~al.}(2019){Litke}, {Marrone}, {Spilker}, {Aravena},
  {B{\'e}thermin}, {Chapman}, {Chen}, {de Breuck}, {Dong}, {Gonzalez}, {Greve},
  {Hayward}, {Hezaveh}, {Jarugula}, {Ma}, {Morningstar}, {Narayanan}, {Phadke},
  {Reuter}, {Vieira}, \& {Weiss}}]{2019ApJ...870...80L}
{Litke}, K.~C., {Marrone}, D.~P., {Spilker}, J.~S., {et~al.} 2019, \apj, 870,
  80, \dodoi{10.3847/1538-4357/aaf057}

\bibitem[{{Lower} {et~al.}(2020){Lower}, {Narayanan}, {Leja}, {Johnson},
  {Conroy}, \& {Dav{\'e}}}]{lower2020ApJ...904...33L}
{Lower}, S., {Narayanan}, D., {Leja}, J., {et~al.} 2020, \apj, 904, 33,
  \dodoi{10.3847/1538-4357/abbfa7}

\bibitem[{{Lower} {et~al.}(2022){Lower}, {Narayanan}, {Leja}, {Johnson},
  {Conroy}, \& {Dav{\'e}}}]{lower_2022_dust_attn}
---. 2022, \apj, 931, 14, \dodoi{10.3847/1538-4357/ac6959}

\bibitem[{{Marrone} {et~al.}(2018){Marrone}, {Spilker}, {Hayward}, {Vieira},
  {Aravena}, {Ashby}, {Bayliss}, {B{\'e}thermin}, {Brodwin}, {Bothwell},
  {Carlstrom}, {Chapman}, {Chen}, {Crawford}, {Cunningham}, {De Breuck},
  {Fassnacht}, {Gonzalez}, {Greve}, {Hezaveh}, {Lacaille}, {Litke}, {Lower},
  {Ma}, {Malkan}, {Miller}, {Morningstar}, {Murphy}, {Narayanan}, {Phadke},
  {Rotermund}, {Sreevani}, {Stalder}, {Stark}, {Strandet}, {Tang}, \&
  {Wei{\ss}}}]{2018Natur.553...51M}
{Marrone}, D.~P., {Spilker}, J.~S., {Hayward}, C.~C., {et~al.} 2018, \nat, 553,
  51, \dodoi{10.1038/nature24629}

\bibitem[{Massey \& Refregier(2005)}]{Massey_2005}
Massey, R., \& Refregier, A. 2005, Monthly Notices of the Royal Astronomical
  Society, 363, 197, \dodoi{10.1111/j.1365-2966.2005.09453.x}

\bibitem[{{McElwain} {et~al.}(2023){McElwain}, {Feinberg}, {Perrin}, {Clampin},
  {Mountain}, {Lallo}, {Lajoie}, {Kimble}, {Bowers}, {Stark}, {Acton},
  {Atkinson}, {Barinek}, {Barto}, {Basinger}, {Beck}, {Bergkoetter}, {Bluth},
  {Boucarut}, {Brady}, {Brooks}, {Brown}, {Byard}, {Carey}, {Carrasquilla},
  {Chae}, {Chaney}, {Chayer}, {Chonis}, {Cohen}, {Cole}, {Comeau}, {Coon},
  {Coppock}, {Coyle}, {Dean}, {Dziak}, {Eisenhower}, {Flagey}, {Franck},
  {Gallagher}, {Gilman}, {Glassman}, {Green}, {Grieco}, {Haase},
  {Hadjimichael}, {Hagopian}, {Hahn}, {Hartig}, {Havey}, {Hayden}, {Hellekson},
  {Hicks}, {Holfeltz}, {Howard}, {Huguet}, {Jahne}, {Johnson}, {Johnston},
  {Jurling}, {Kegley}, {Kennard}, {Keski-Kuha}, {Knight}, {Kulp}, {Levi},
  {Levine}, {Lightsey}, {Luetgens}, {Mather}, {Matthews}, {McKay}, {Mehalick},
  {Mel{\'e}ndez}, {Mosier}, {Murphy}, {Nelan}, {Niedner}, {Nol}, {Ohara},
  {Ohl}, {Olczak}, {Osborne}, {Park}, {Perrygo}, {Pueyo}, {Redding}, {Regan},
  {Reynolds}, {Rifelli}, {Rigby}, {Sabatke}, {Saif}, {Scorse}, {Seo}, {Shi},
  {Sigrist}, {Smith}, {Smith}, {Smith}, {Sohn}, {Stahl}, {Telfer}, {Terlecki},
  {Texter}, {Van Buren}, {Van Campen}, {Vila}, {Voyton}, {Waldman}, {Walker},
  {Weiser}, {Wells}, {West}, {Whitman}, {Wolf}, \& {Zielinski}}]{McElwain_2023}
{McElwain}, M.~W., {Feinberg}, L.~D., {Perrin}, M.~D., {et~al.} 2023, \pasp,
  135, 058001, \dodoi{10.1088/1538-3873/acada0}

\bibitem[{{McMullin} {et~al.}(2007){McMullin}, {Waters}, {Schiebel}, {Young},
  \& {Golap}}]{mcmullin2007ASPC..376..127M}
{McMullin}, J.~P., {Waters}, B., {Schiebel}, D., {Young}, W., \& {Golap}, K.
  2007, in Astronomical Society of the Pacific Conference Series, Vol. 376,
  Astronomical Data Analysis Software and Systems XVI, ed. R.~A. {Shaw},
  F.~{Hill}, \& D.~J. {Bell}, 127

\bibitem[{{Micha{\l}owski} {et~al.}(2014){Micha{\l}owski}, {Hayward}, {Dunlop},
  {Bruce}, {Cirasuolo}, {Cullen}, \&
  {Hernquist}}]{michalowski2014A&A...571A..75M}
{Micha{\l}owski}, M.~J., {Hayward}, C.~C., {Dunlop}, J.~S., {et~al.} 2014,
  \aap, 571, A75, \dodoi{10.1051/0004-6361/201424174}

\bibitem[{{Mobasher} {et~al.}(2015){Mobasher}, {Dahlen}, {Ferguson},
  {Acquaviva}, {Barro}, {Finkelstein}, {Fontana}, {Gruetzbauch}, {Johnson},
  {Lu}, {Papovich}, {Pforr}, {Salvato}, {Somerville}, {Wiklind}, {Wuyts},
  {Ashby}, {Bell}, {Conselice}, {Dickinson}, {Faber}, {Fazio}, {Finlator},
  {Galametz}, {Gawiser}, {Giavalisco}, {Grazian}, {Grogin}, {Guo}, {Hathi},
  {Kocevski}, {Koekemoer}, {Koo}, {Newman}, {Reddy}, {Santini}, \&
  {Wechsler}}]{mobasher2015ApJ...808..101M}
{Mobasher}, B., {Dahlen}, T., {Ferguson}, H.~C., {et~al.} 2015, \apj, 808, 101,
  \dodoi{10.1088/0004-637X/808/1/101}

\bibitem[{{Nelson} {et~al.}(2022){Nelson}, {Suess}, {Bezanson}, {Price}, {van
  Dokkum}, {Leja}, {Wang}, {Whitaker}, {Labb{\'e}}, {Barrufet}, {Brammer},
  {Eisenstein}, {Heintz}, {Johnson}, {Mathews}, {Miller}, {Oesch}, {Sandles},
  {Setton}, {Speagle}, {Tacchella}, {Tadaki}, \&
  {Weaver}}]{nelson2022arXiv220801630N}
{Nelson}, E.~J., {Suess}, K.~A., {Bezanson}, R., {et~al.} 2022, arXiv e-prints,
  arXiv:2208.01630, \dodoi{10.48550/arXiv.2208.01630}

\bibitem[{{Peng} {et~al.}(2022){Peng}, {Vishwas}, {Stacey}, {Nikola},
  {Lamarche}, {Rooney}, {Ball}, {Ferkinhoff}, \&
  {Spoon}}]{peng2022arXiv221016968P}
{Peng}, B., {Vishwas}, A., {Stacey}, G., {et~al.} 2022, arXiv e-prints,
  arXiv:2210.16968.
\newblock \doarXiv{2210.16968}

\bibitem[{{Peng} {et~al.}(2002){Peng}, {Ho}, {Impey}, \&
  {Rix}}]{peng2002AJ....124..266P}
{Peng}, C.~Y., {Ho}, L.~C., {Impey}, C.~D., \& {Rix}, H.-W. 2002, \aj, 124,
  266, \dodoi{10.1086/340952}

\bibitem[{{Peng} {et~al.}(2010){Peng}, {Ho}, {Impey}, \&
  {Rix}}]{peng2010AJ....139.2097P}
---. 2010, \aj, 139, 2097, \dodoi{10.1088/0004-6256/139/6/2097}

\bibitem[{{Perrin} {et~al.}(2014){Perrin}, {Sivaramakrishnan}, {Lajoie},
  {Elliott}, {Pueyo}, {Ravindranath}, \& {Albert}}]{perrin2014}
{Perrin}, M.~D., {Sivaramakrishnan}, A., {Lajoie}, C.-P., {et~al.} 2014, in
  Society of Photo-Optical Instrumentation Engineers (SPIE) Conference Series,
  Vol. 9143, Space Telescopes and Instrumentation 2014: Optical, Infrared, and
  Millimeter Wave, ed. J.~{Oschmann}, Jacobus~M., M.~{Clampin}, G.~G. {Fazio},
  \& H.~A. {MacEwen}, 91433X, \dodoi{10.1117/12.2056689}

\bibitem[{{Perrin} {et~al.}(2012){Perrin}, {Soummer}, {Elliott}, {Lallo}, \&
  {Sivaramakrishnan}}]{perrin2012}
{Perrin}, M.~D., {Soummer}, R., {Elliott}, E.~M., {Lallo}, M.~D., \&
  {Sivaramakrishnan}, A. 2012, in Society of Photo-Optical Instrumentation
  Engineers (SPIE) Conference Series, Vol. 8442, Space Telescopes and
  Instrumentation 2012: Optical, Infrared, and Millimeter Wave, ed. M.~C.
  {Clampin}, G.~G. {Fazio}, H.~A. {MacEwen}, \& J.~{Oschmann}, Jacobus~M.,
  84423D, \dodoi{10.1117/12.925230}

\bibitem[{{Perry} {et~al.}(2022){Perry}, {Chapman}, {Smail}, \&
  {Bertoldi}}]{perry2022arXiv221008191P}
{Perry}, R.~W., {Chapman}, S.~C., {Smail}, I., \& {Bertoldi}, F. 2022, arXiv
  e-prints, arXiv:2210.08191, \dodoi{10.48550/arXiv.2210.08191}

\bibitem[{{Pillepich} {et~al.}(2019){Pillepich}, {Nelson}, {Springel},
  {Pakmor}, {Torrey}, {Weinberger}, {Vogelsberger}, {Marinacci}, {Genel}, {van
  der Wel}, \& {Hernquist}}]{pillepich2019MNRAS.490.3196P}
{Pillepich}, A., {Nelson}, D., {Springel}, V., {et~al.} 2019, \mnras, 490,
  3196, \dodoi{10.1093/mnras/stz2338}

\bibitem[{{Planck Collaboration} {et~al.}(2020){Planck Collaboration},
  {Aghanim}, {Akrami}, {Ashdown}, {Aumont}, {Baccigalupi}, {Ballardini},
  {Banday}, {Barreiro}, {Bartolo}, {Basak}, {Battye}, {Benabed}, {Bernard},
  {Bersanelli}, {Bielewicz}, {Bock}, {Bond}, {Borrill}, {Bouchet}, {Boulanger},
  {Bucher}, {Burigana}, {Butler}, {Calabrese}, {Cardoso}, {Carron},
  {Challinor}, {Chiang}, {Chluba}, {Colombo}, {Combet}, {Contreras}, {Crill},
  {Cuttaia}, {de Bernardis}, {de Zotti}, {Delabrouille}, {Delouis}, {Di
  Valentino}, {Diego}, {Dor{\'e}}, {Douspis}, {Ducout}, {Dupac}, {Dusini},
  {Efstathiou}, {Elsner}, {En{\ss}lin}, {Eriksen}, {Fantaye}, {Farhang},
  {Fergusson}, {Fernandez-Cobos}, {Finelli}, {Forastieri}, {Frailis},
  {Fraisse}, {Franceschi}, {Frolov}, {Galeotta}, {Galli}, {Ganga},
  {G{\'e}nova-Santos}, {Gerbino}, {Ghosh}, {Gonz{\'a}lez-Nuevo}, {G{\'o}rski},
  {Gratton}, {Gruppuso}, {Gudmundsson}, {Hamann}, {Handley}, {Hansen},
  {Herranz}, {Hildebrandt}, {Hivon}, {Huang}, {Jaffe}, {Jones}, {Karakci},
  {Keih{\"a}nen}, {Keskitalo}, {Kiiveri}, {Kim}, {Kisner}, {Knox},
  {Krachmalnicoff}, {Kunz}, {Kurki-Suonio}, {Lagache}, {Lamarre}, {Lasenby},
  {Lattanzi}, {Lawrence}, {Le Jeune}, {Lemos}, {Lesgourgues}, {Levrier},
  {Lewis}, {Liguori}, {Lilje}, {Lilley}, {Lindholm}, {L{\'o}pez-Caniego},
  {Lubin}, {Ma}, {Mac{\'\i}as-P{\'e}rez}, {Maggio}, {Maino}, {Mandolesi},
  {Mangilli}, {Marcos-Caballero}, {Maris}, {Martin}, {Martinelli},
  {Mart{\'\i}nez-Gonz{\'a}lez}, {Matarrese}, {Mauri}, {McEwen}, {Meinhold},
  {Melchiorri}, {Mennella}, {Migliaccio}, {Millea}, {Mitra},
  {Miville-Desch{\^e}nes}, {Molinari}, {Montier}, {Morgante}, {Moss}, {Natoli},
  {N{\o}rgaard-Nielsen}, {Pagano}, {Paoletti}, {Partridge}, {Patanchon},
  {Peiris}, {Perrotta}, {Pettorino}, {Piacentini}, {Polastri}, {Polenta},
  {Puget}, {Rachen}, {Reinecke}, {Remazeilles}, {Renzi}, {Rocha}, {Rosset},
  {Roudier}, {Rubi{\~n}o-Mart{\'\i}n}, {Ruiz-Granados}, {Salvati}, {Sandri},
  {Savelainen}, {Scott}, {Shellard}, {Sirignano}, {Sirri}, {Spencer},
  {Sunyaev}, {Suur-Uski}, {Tauber}, {Tavagnacco}, {Tenti}, {Toffolatti},
  {Tomasi}, {Trombetti}, {Valenziano}, {Valiviita}, {Van Tent}, {Vibert},
  {Vielva}, {Villa}, {Vittorio}, {Wandelt}, {Wehus}, {White}, {White},
  {Zacchei}, \& {Zonca}}]{planck2020A&A...641A...6P}
{Planck Collaboration}, {Aghanim}, N., {Akrami}, Y., {et~al.} 2020, \aap, 641,
  A6, \dodoi{10.1051/0004-6361/201833910}

\bibitem[{Refregier(2003)}]{Refregier_2003}
Refregier, A. 2003, Monthly Notices of the Royal Astronomical Society, 338, 35,
  \dodoi{10.1046/j.1365-8711.2003.05901.x}

\bibitem[{Refregier \& Bacon(2003)}]{Refregier_2003b}
Refregier, A., \& Bacon, D. 2003, Monthly Notices of the Royal Astronomical
  Society, 338, 48, \dodoi{10.1046/j.1365-8711.2003.05902.x}

\bibitem[{{Reuter} {et~al.}(2020){Reuter}, {Vieira}, {Spilker}, {Weiss},
  {Aravena}, {Archipley}, {B{\'e}thermin}, {Chapman}, {De Breuck}, {Dong},
  {Everett}, {Fu}, {Greve}, {Hayward}, {Hill}, {Hezaveh}, {Jarugula}, {Litke},
  {Malkan}, {Marrone}, {Narayanan}, {Phadke}, {Stark}, \&
  {Strandet}}]{reuter2020ApJ...902...78R}
{Reuter}, C., {Vieira}, J.~D., {Spilker}, J.~S., {et~al.} 2020, \apj, 902, 78,
  \dodoi{10.3847/1538-4357/abb599}

\bibitem[{Rigby {et~al.}(2023)Rigby, Perrin, McElwain, Kimble, Friedman, Lallo,
  Doyon, Feinberg, Ferruit, Glasse, Rieke, Rieke, Wright, Willott, Colon,
  Milam, Neff, Stark, Valenti, Abell, Abney, Abul-Huda, Acton, Adams, Adler,
  Aguilar, Ahmed, Albert, Alberts, Aldridge, Allen, Altenburg,
  {\'{A}}lvarez-M{\'{a}}rquez, de~Oliveira, Andersen, Anderson, Anderson,
  Argyriou, Armstrong, Arribas, Artigau, Arvai, Atkinson, Bacon, Bair, Banks,
  Barrientes, Barringer, Bartosik, Bast, Baudoz, Beatty, Bechtold, Beck,
  Bergeron, Bergkoetter, Bhatawdekar, Birkmann, Blazek, Blome, Boccaletti,
  Böker, Boia, Bonaventura, Bond, Bosley, Boucarut, Bourque, Bouwman, Bower,
  Bowers, Boyer, Bradley, Brady, Braun, Breda, Bresnahan, Bright, Britt,
  Bromenschenkel, Brooks, Brooks, Brown, Brown, Brown, Bunker, Burger,
  Bushouse, Cale, Cameron, Cameron, Canipe, Caplinger, Caputo, Cara, Carey,
  Carniani, Carrasquilla, Carruthers, Case, Catherine, Chance, Chapman,
  Charlot, Charlow, Chayer, Chen, Cherinka, Chichester, Chilton, Chonis,
  Clampin, Clark, Clark, Coe, Coleman, Comber, Comeau, Connolly, Cooper,
  Cooper, Coppock, Correnti, Cossou, Coulais, Coyle, Cracraft, Curti, Cuturic,
  Davis, Davis, Dean, DeLisa, deMeester, Dencheva, Dencheva, DePasquale,
  Deschenes, Örs Hunor~Detre, Diaz, Dicken, DiFelice, Dillman, Dixon, Doggett,
  Donaldson, Douglas, DuPrie, Dupuis, Durning, Easmin, Eck, Edeani, Egami,
  Ehrenwinkler, Eisenhamer, Eisenhower, Elie, Elliott, Elliott, Ellis,
  Engesser, Espinoza, Etienne, Etxaluze, Falini, Feeney, Ferry, Filippazzo,
  Fincham, Fix, Flagey, Florian, Flynn, Fontanella, Ford, Forshay, Fox, Franz,
  Fu, Fullerton, Galkin, Galyer, Mar{\'{\i}}n, Gardner, Gardner, Garland,
  Garrett, Gasman, Gaspar, Gaudreau, Gauthier, Geers, Geithner, Gennaro,
  Giardino, Girard, Giuliano, Glassmire, Glauser, Glazer, Godfrey, Golimowski,
  Gollnitz, Gong, Gonzaga, Gordon, Gordon, Goudfrooij, Greene, Greenhouse,
  Grimaldi, Groebner, Grundy, Guillard, Gutman, Ha, Haderlein, Hagedorn,
  Hainline, Haley, Hami, Hamilton, Hammel, Hansen, Harkins, Harr, Hart, Hart,
  Hartig, Hashimoto, Haskins, Hathaway, Havey, Hayden, Hecht, Heller-Boyer,
  Henriques, Henry, Hermann, Hernandez, Hesman, Hicks, Hilbert, Hines, Hoffman,
  Holfeltz, Holler, Hoppa, Hott, Howard, Howard, Hunter, Hunter, Hurst,
  Husemann, Hustak, Ignat, Illingworth, Irish, Jackson, Jahromi, Jakobsen,
  James, James, Januszewski, Jenkins, Jirdeh, Johnson, Johnson, Jones, Jones,
  Jones, Jones, Jordan, Jordan, Jurczyk, Jurling, Kaleida, Kalmanson, Kammerer,
  Kang, Kao, Karakla, Kavanagh, Kelly, Kendrew, Kennedy, Kenny, Keski-kuha,
  Keyes, Kidwell, Kinzel, Kirk, Kirkpatrick, Kirshenblat, Klaassen, Knapp,
  Knight, Knollenberg, Koehler, Koekemoer, Kovacs, Kulp, Kumari, Kyprianou,
  Massa, Labador, Labiano, Lagage, Lajoie, Lallo, Lam, Lamb, Lambros,
  Lampenfield, Langston, Larson, Law, Lawrence, Lee, Leisenring, Lepo,
  Leveille, Levenson, Levine, Levy, Lewis, Lewis, Libralato, Lightsey, Link,
  Liu, Lo, Lockwood, Logue, Long, Long, Loomis, Lopez-Caniego, Alvarez,
  Love-Pruitt, Lucy, Luetzgendorf, Maghami, Maiolino, Major, Malla, Malumuth,
  Manjavacas, Mannfolk, Marrione, Marston, Martel, Maschmann, Masci,
  Masciarelli, Maszkiewicz, Mather, McKenzie, McLean, McMaster, Melbourne,
  Mel{\'{e}}ndez, Menzel, Merz, Meyett, Meza, Miskey, Misselt, Moller,
  Morrison, Morse, Moseley, Mosier, Mountain, Mueckay, Mueller, Mullally,
  Murphy, Murray, Murray, Mustelier, Muzerolle, Mycroft, Myers, Myrick,
  Nanavati, Nance, Nayak, Naylor, Nelan, Nickson, Nielson, Nieto-Santisteban,
  Nikolov, Noriega-Crespo, O'Shaughnessy, O'Sullivan, Ochs, Ogle, Oleszczuk,
  Olmsted, Osborne, Ottens, Owens, Pacifici, Pagan, Page, Park, Parrish,
  Patapis, Paul, Pauly, Pavlovsky, Pedder, Peek, Pena-Guerrero, Penanen, Perez,
  Perna, Perriello, Phillips, Pietraszkiewicz, Pinaud, Pirzkal, Pitman,
  Piwowar, Platais, Player, Plesha, Pollizi, Polster, Pontoppidan, Porterfield,
  Proffitt, Pueyo, Pulliam, Quirt, Neira, Alarcon, Ramsay, Rapp, Rapp,
  Rauscher, Ravindranath, Rawle, Regan, Reichard, Reis, Ressler, Rest,
  Reynolds, Rhue, Richon, Rickman, Ridgaway, Ritchie, Rix, Robberto, Robinson,
  Robinson, Robinson, Rock, Rodriguez, Pino, Roellig, Rohrbach, Roman,
  Romelfanger, Rose, Roteliuk, Roth, Rothwell, Rowlands, Roy, Royer, Royle,
  Rui, Rumler, Runnels, Russ, Rustamkulov, Ryden, Ryer, Sabata, Sabatke, Sabbi,
  Samuelson, Sapp, Sappington, Sargent, Sauer, Scheithauer, Schlawin, Schlitz,
  Schmitz, Schneider, Schreiber, Schulze, Schwab, Scott, Sembach, Shanahan,
  Shaughnessy, Shaw, Shawger, Shay, Sheehan, Shen, Sherman, Shiao, Shih,
  Shivaei, Sienkiewicz, Sing, Sirianni, Sivaramakrishnan, Skipper, Sloan,
  Slocum, Slowinski, Smith, Smith, Smith, Smith, Snyder, Soh, Sohn, Soto,
  Spencer, Stallcup, Stansberry, Starr, Starr, Stewart, Stiavelli, Straughn,
  Strickland, Stys, Summers, Sun, Sunnquist, Swade, Swam, Swaters, Swoish,
  Taylor, Taylor, Plate, Tea, Teague, Telfer, Temim, Thatte, Thompson,
  Thompson, Thomson, Tikkanen, Tippet, Todd, Toolan, Tran, Trejo, Truong,
  Tsukamoto, Tustain, Tyra, Ubeda, Underwood, Uzzo, Campen, Vandal,
  Vandenbussche, Vila, Volk, Wahlgren, Waldman, Walker, Wander, Warfield,
  Warner, Wasiak, Watkins, Weaver, Weilert, Weiser, Weiss, Weissman, Welty,
  West, Wheate, Wheatley, Wheeler, White, Whiteaker, Whitehouse, Whiteleather,
  Whitman, Williams, Willmer, Willoughby, Wilson, Wirth, Wislowski, Wolf,
  Wolfe, Wolff, Workman, Wright, Wu, Wu, Wymer, Yates, Yeager, Yeates, Yerger,
  Yoon, Young, Yu, Zak, Zeidler, Zhou, Zielinski, Zincke, \&
  Zonak}]{Rigby_2023}
Rigby, J., Perrin, M., McElwain, M., {et~al.} 2023, Publications of the
  Astronomical Society of the Pacific, 135, 048001,
  \dodoi{10.1088/1538-3873/acb293}

\bibitem[{{Rizzo} {et~al.}(2020){Rizzo}, {Vegetti}, {Powell}, {Fraternali},
  {McKean}, {Stacey}, \& {White}}]{rizzo2020Natur.584..201R}
{Rizzo}, F., {Vegetti}, S., {Powell}, D., {et~al.} 2020, \nat, 584, 201,
  \dodoi{10.1038/s41586-020-2572-6}

\bibitem[{{Robertson} {et~al.}(2006){Robertson}, {Bullock}, {Cox}, {Di Matteo},
  {Hernquist}, {Springel}, \& {Yoshida}}]{robertson2006ApJ...645..986R}
{Robertson}, B., {Bullock}, J.~S., {Cox}, T.~J., {et~al.} 2006, \apj, 645, 986,
  \dodoi{10.1086/504412}

\bibitem[{{Roman-Oliveira} {et~al.}(2023){Roman-Oliveira}, {Fraternali}, \&
  {Rizzo}}]{romanoliveira2023arXiv230203049R}
{Roman-Oliveira}, F., {Fraternali}, F., \& {Rizzo}, F. 2023, arXiv e-prints,
  arXiv:2302.03049, \dodoi{10.48550/arXiv.2302.03049}

\bibitem[{{Sanders} {et~al.}(1988){Sanders}, {Soifer}, {Elias}, {Madore},
  {Matthews}, {Neugebauer}, \& {Scoville}}]{sanders1988ApJ...325...74S}
{Sanders}, D.~B., {Soifer}, B.~T., {Elias}, J.~H., {et~al.} 1988, \apj, 325,
  74, \dodoi{10.1086/165983}

\bibitem[{{Sotillo-Ramos} {et~al.}(2022){Sotillo-Ramos}, {Pillepich},
  {Donnari}, {Nelson}, {Eisert}, {Rodriguez-Gomez}, {Joshi}, {Vogelsberger}, \&
  {Hernquist}}]{sotilloramos2022MNRAS.516.5404S}
{Sotillo-Ramos}, D., {Pillepich}, A., {Donnari}, M., {et~al.} 2022, \mnras,
  516, 5404, \dodoi{10.1093/mnras/stac2586}

\bibitem[{{Spilker} {et~al.}(2020){Spilker}, {Aravena}, {Phadke},
  {B{\'e}thermin}, {Chapman}, {Dong}, {Gonzalez}, {Hayward}, {Hezaveh},
  {Litke}, {Malkan}, {Marrone}, {Narayanan}, {Reuter}, {Vieira}, \&
  {Wei{\ss}}}]{spilker2020ApJ...905...86S}
{Spilker}, J.~S., {Aravena}, M., {Phadke}, K.~A., {et~al.} 2020, \apj, 905, 86,
  \dodoi{10.3847/1538-4357/abc4e6}

\bibitem[{{Steinmetz} \& {Navarro}(2002)}]{steinmetz2002NewA....7..155S}
{Steinmetz}, M., \& {Navarro}, J.~F. 2002, \na, 7, 155,
  \dodoi{10.1016/S1384-1076(02)00102-1}

\bibitem[{{THE CASA TEAM} {et~al.}(2022){THE CASA TEAM}, {Bean}, {Bhatnagar},
  {Castro}, {Donovan Meyer}, {Emonts}, {Garcia}, {Garwood}, {Golap}, {Gonzalez
  Villalba}, {Harris}, {Hayashi}, {Hoskins}, {Hsieh}, {Jagannathan},
  {Kawasaki}, {Keimpema}, {Kettenis}, {Lopez}, {Marvil}, {Masters},
  {McNichols}, {Mehringer}, {Miel}, {Moellenbrock}, {Montesino}, {Nakazato},
  {Ott}, {Petry}, {Pokorny}, {Raba}, {Rau}, {Schiebel}, {Schweighart},
  {Sekhar}, {Shimada}, {Small}, {Steeb}, {Sugimoto}, {Suoranta}, {Tsutsumi},
  {van Bemmel}, {Verkouter}, {Wells}, {Xiong}, {Szomoru}, {Griffith},
  {Glendenning}, \& {Kern}}]{casa2022}
{THE CASA TEAM}, {Bean}, B., {Bhatnagar}, S., {et~al.} 2022, arXiv e-prints,
  arXiv:2210.02276.
\newblock \doarXiv{2210.02276}

\bibitem[{{Vieira} {et~al.}(2013){Vieira}, {Marrone}, {Chapman}, {De Breuck},
  {Hezaveh}, {Wei{\ensuremath{\beta}}}, {Aguirre}, {Aird}, {Aravena}, {Ashby},
  {Bayliss}, {Benson}, {Biggs}, {Bleem}, {Bock}, {Bothwell}, {Bradford},
  {Brodwin}, {Carlstrom}, {Chang}, {Crawford}, {Crites}, {de Haan}, {Dobbs},
  {Fomalont}, {Fassnacht}, {George}, {Gladders}, {Gonzalez}, {Greve},
  {Gullberg}, {Halverson}, {High}, {Holder}, {Holzapfel}, {Hoover}, {Hrubes},
  {Hunter}, {Keisler}, {Lee}, {Leitch}, {Lueker}, {Luong-van}, {Malkan},
  {McIntyre}, {McMahon}, {Mehl}, {Menten}, {Meyer}, {Mocanu}, {Murphy},
  {Natoli}, {Padin}, {Plagge}, {Reichardt}, {Rest}, {Ruel}, {Ruhl}, {Sharon},
  {Schaffer}, {Shaw}, {Shirokoff}, {Spilker}, {Stalder}, {Staniszewski},
  {Stark}, {Story}, {Vanderlinde}, {Welikala}, \&
  {Williamson}}]{vieira2013Natur.495..344V}
{Vieira}, J.~D., {Marrone}, D.~P., {Chapman}, S.~C., {et~al.} 2013, \nat, 495,
  344, \dodoi{10.1038/nature12001}

\end{thebibliography}
\bibliographystyle{aasjournal}

\end{document}